# A systematic review of physical-digital play technology and developmentally relevant child behaviour

Pablo E. Torres*, Philip I. N. Ulrich, Veronica Cucuiat, Mutlu Cukurova, María Fercovic De la Presa, Rose Luckin, Amanda Carr, Thomas Dylan, Abigail Durrant, John Vines, & Shaun Lawson.

## Research Highlights

- We reviewed how physical-digital play engaged children in developmentally relevant behaviour

- 31 relevant articles were identified; 17 higher-quality articles were included for synthesis

- Physical-digital play promoted behaviours relevant for self-monitoring, collaboration, decision-making, problem-solving and physical activity

- The review identified specific ways in which these skills were promoted by play interactivity

- A theoretical framework (the goal-oriented tool-mediated action framework) is developed to explain the behavioural affordances found

## Abstract

New interactive physical-digital play technologies are shaping the way children play. These technologies refer to digital play technologies that engage children in analogue (non-digital) forms of behaviour, either alone or with others. Current interactive physical-digital play technologies include robots, digital agents, mixed or augmented reality devices, and smart-eye based gaming. Little is known, however, about the ways in which these technologies could promote or damage child development. This systematic review was aimed at understanding if and how these physical-digital play technologies promoted developmentally relevant behaviour (related to transferable skills and physical activity) in typically developing 0 to 12 year-olds. Psychology, Education, and Computer Science databases were searched producing 635 papers. A total of 31 papers met the inclusion criteria, of which 17 were of high enough quality to be included for synthesis. A theoretical framework was developed to guide our review and a thematic analysis was applied to find patterns across empirical studies. Results indicate that these new interactive play technologies could have a positive effect on children's developmentally relevant behaviour. The review identified specific ways in which different behaviours were promoted by the play interactivity. Providing information about own performance promoted self-monitoring. Slowing interactivity, play interdependency, and joint object accessibility promoted collaboration. Offering delimited choices promoted decision making. Problem solving and physical activity were promoted by requiring children to engage in them to keep playing. Four overarching principles underpinned the ways in which phygital play technologies afforded child behaviour. These included social expectations framing play situations, the directiveness of action regulations (i.e., inviting, guiding or forcing behaviours), the technical features of play technologies (digital play mechanics and physical characteristics), and the alignment between play goals, play technology and the play behaviours promoted.

Keywords: Systematic review, digital play, child development, child behaviour, child-computer interactions

*Corresponding author. Faculty of Education, University of Cambridge. Email: pelt2@cam.ac.uk ; torresp.uk@gmail.com

# 1 Introduction

A new generation of physical–digital ('phygital') play technologies [1, 2, 3] is appearing in the form of smart toys and digitally augmented play spaces, and making its way into children's lives [4]. These new technologies come to complement a mature video-game industry that has long raised controversy in terms of its benefits for child development. Although video-game studies have identified positive effects to children's learning and development [5, 6, 7, 8], negative consequences have dominated the rhetoric. Studies have consistently linked the enormous recent increase in the children's time spent playing video-games [9, 10, 11] to a reduction in outdoor play and physical activity, lower psychosocial wellbeing or decreased attention-span [12, 13, 14, 15, 16]. This has generated an interest to explore how technologies may be leveraged to counter some of these negative impacts. To do this, designers are creating interactive digital experiences that mirror traditional play contexts to increase the physicality and social interactivity of digital play [1]. This novel and mostly unexplored area of literature is the focus of the present review.

The review focuses on 'physical–digital (phygital) play', defined here as digital play that also engages children in analogue (non-digital) forms of behaviour, either alone or with others. Some phygital play devices mix and sync screens with tangible objects [2, 17] and bring small figurine toys-to-life in the virtual world [3]. Others use screen-free digitally enhanced objects to engage children in more organic play behaviours embedded in the physical world, such as in playgrounds or forests [18, 19]. These types of objects generally trigger sensory and/or verbal feedback from toys and play spaces which children might use to play in the physical realm in more open-ended ways [20]. However, the analogue behaviours promoted by phygital technologies examined in relation to child development to date are, with the exception of robots, still largely reliant on screen-based interactions with or between players (bodily controlled video-games, virtual agents, exergames, and augmented reality).

Previous studies have found mixed and therefore inconclusive evidence of the effects of screen technology use for child learning and development. The studies show both benefits [5, 6, 7, 8] and detriments [12, 21, 15, 13, 16] of such a type of technology for children's cognitive skills, social skills and activity, physical activity, content knowledge and general play behaviour. New research has started to disentangle these inconsistent effects. The answer seems to be in the interactive engagements triggered by technology. Specifically, higher levels of contingent interactive digital responses to children's behaviours seem to promote learning and development. For example, joint media engagements have been found to lead to higher levels of learning achievement than when media is used individually [22]. Children can learn as much from video-chats as from in person interactions, and either of these is better for learning than one-way (recorded, non-interactive) video demonstrations [23].

The relevance of engaging interactivity seems to also be highlighted by recent reviews on digital play technologies. Based on 14 studies, a review by Güneş [24] concluded that digital game-based learning has positive effects on children's learning of specific contents or skills (e.g., motor skills, logics of programming); a result also found for higher-education by Subhash & Cudney [25] systematic review of 41 studies. Most interestingly, Güneş [24] review shows that screen play technologies can have a positive effect on child learning and development when they promote active rather than passive engagement. A similar conclusion was achieved in the systematic review of 17 studies conducted by Yanti, Rosmansyah, & Dabarsyah [26]. The authors found that interactivity was the main vehicle used by serious games to promote learning. This is a result also seconded by Griffith, Hagan, Heymann, Helfim, & Bagner [27] systematic review of 35 studies about the effect of play apps on children's literacy and mathematic skills. However, despite overall positive effects of the interactivity of digital play, not all reviews converge on equating digital interactivity to developmental dividends among children. In particular, and although only based on five studies, the systematic review by Bochicchio et al. [28] indicates that interactive digital play can be both good and bad for children. It can increase children's pro-social and anti-social behaviours, facilitate negotiations processes but also isolation, and generate cooperation as well as competition.

There have been mixed academic views about whether engagement with new types of phygital smart toys and games is positive or negative for children. On the one hand, Jenkins et al. [29] considers that the skills involved in using these new types of technologically augmented play materials are the skills that people will need in the future. Similarly, Kafai [30] considers that children experience positive feedback and feelings of control with these technologies, making it beneficial for their motivation to keep learning and improving. However, researchers have also acknowledged that phygital play could have negative consequences for children [30, 31]. Evidence from studies of these types of new technologies seems to be provisional and scarce. As expressed by Bergen, Davis and Abbitt [32] when talking about changes in technology-augmentation in traditional and innovative play materials, "Whether these differences [changes] in the play experiences of children and adolescents will result in differences in brain development and subsequently in social, emotional, moral, and cognitive development is presently only at the level of speculation" (p.47).

# 2  Aim of the review

This review aims to help start addressing the dearth of evidence and resulting speculation that Bergen, Davis, and Abbitt [32] refer to. Recent reviews and studies presented above have been instrumental to uncover the interactivity link between digital technologies or digital play and child development. However, these have not scrutinised *how* is that the interactivity of play technologies promotes child learning or development in practice. The present review aims to advance such understanding. It aims to do this by synthesising literature on these new forms of physical–digital play technologies. Following Eisenberg [33] our effort is not only descriptive but theoretical. In particular, we expect the results from this review help advance a conceptual framework about how interactive phygital play technology promotes developmentally relevant child behaviour. Two research questions guided our systematic review process:

RQ.1 What type of social, emotional, cognitive or physical developmentally relevant actions do phygital play technologies afford for children?

RQ.2 How are these affordances delivered by phygital play technologies?

The first question was aimed at identifying the different types of developmentally relevant behavioural partnerships between phygital play technologies and children's play activity. The second question focused on identifying the technological features through which such partnerships are facilitated.

When defining developmentally relevant behaviour, we considered any type of behavioural activity studied within developmental or educational psychology. This included any actions (and indicators) related to practice of cognition, emotions, social competences, and physical activity. When defining the scope of play we followed part of Smith and Pellegrini's [34] characterisations of play, who conceptualise it as an activity carried out for its own sake and enjoyment rather than a productive aim. In the next section we elaborate further on our working definition of play, our general understanding of the nature of development during childhood, as well as the goal-oriented tool-mediated action framework guiding the current review.

# 3 Theoretical framework

## 3.1 A working definition of play

Play can be defined in many different ways. As indicated by Zosh et al. [35] in their review of the concept, 'play' could be considered to cover a complex and wide spectrum of human (or animal) activity. The review cites various definitions of play. Most of the theorists it visits understand play as an activity done for its own sake (Garvey; Gray; Piaget; Smith and Pellegrini; Stuart Brown) which includes experiencing enjoyment, pleasure or a stress-free mind (Garvey; Gray; Smith & Pellegrini; Weisberg et al.). Some reviewed authors also add that play should be defined as a flexible (Stuart Brown, Smith and Pellegrini) and child-led activity (Gray; Weisberg et al.). The extent to which play is conceptualised as adding flexibility and being child-led, however, has direct consequences for whether or not activities such as games, guided-play or any other type of structured playful activity could be understood to be play [34]. Considering the multiplicity of perspectives about what play is, and following Zosh et al. [35], it could be argued that a suitable working definition of play should incorporate facets from extant play definitions that are functional and consistent with the specific practical aims of each field of study.

A working definition of play to be used for the improvement of developmentally beneficial play technologies needs to consider the overlaps between technology use and human development. Technology used by humans tends to structure human activity [36, 37] and human thinking [38]. This is relevant for the role that technology can have for child development because the latter tends to be accelerated when scaffolded (structured) towards higher individual functioning [39, 40, 41]. Therefore, we argue that, when studying play technologies in relation to child development, a definition of play allowing structured activities, including closed-ended ones, to be considered play is most productive (see [35] for a similar take for the case of play and learning). In addition, the fact that play could be considered to be an activity carried out for its own sake, differentiates it from a plethora of play-related activities that have started to be researched in recent decades. These include productive activities such as playful learning, creativity as playing, or innovation as playing, carried out motivated by the ends rather than the means of playful activity [42, 35]. This distinction between intrinsic and extrinsic value of play also allows to differentiate play from activities that people also enjoy but that are usually directed by a deliberate productive aim and for which people use analogue or digital technology to enhance their performance rather than to enjoy [43, 44], such as a hobby, volunteering or even working. We, therefore, consider play to be a flow of close- or open-ended activities – a series of connected events  (Sutton-Smith, as cited in Eberle [45]) – that children carry out for their own sake and personal enjoyment rather than with a productive aim.

Such a definition of play is not only theoretically appropriate but also in keeping with the current state of the field of phygital play. As can be seen from current work in the child–computer interaction community, today's phygital play technologies rely on a pre-programmed and expected range of feedback loops of human-technology inputs and responses [1, 2, 3]. Therefore, these technologies lend themselves more for the closed-ended than the open-ended side of the play spectrum. Additionally, rules are at the very heart of the play mechanics of a significant amount of phygital play technologies [46, 47]. This inevitably constrains children to play within the limits of such pre-defined rules [48] rather than in flexible or open-ended ways. Furthermore, even the most embodied current phygital technologies, such as head-up games, tend to still be designed with an adult-determined (learning or recreational) objective in mind [49, 50]. This continues to be the case despite the fact that participatory design methods including children in the design of play technologies have become mainstream within the child–computer interaction community [51].

All of the above does not mean that we think technology could not be used in free and flexible ways. Indeed, in keeping with Heidegger's concept about the readiness-to-hand of human-object / human-world interaction [52], some play technologies are designed to be used in open-ended and flexible ways by letting children impose their own meanings on them whilst playing with them. This can be seen when technology supports pretend play, free musical play, or the creation of own games or stories [53, 54, 55, 56]. However, narrowing our understanding of play technologies only to those supporting more open-ended play would limit a review on current phygital play technologies and child development both empirically (due to the wide range of current technologies) and theoretically (due to the structuring nature of technology use). Consequently, we believe that defining play as an activity carried out for its own sake and enjoyment, and not restricting it to features such as its flexibility or how it lends itself for child-led actions, might allow for a more inclusive and productive analysis of how current phygital play technologies promote developmentally relevant child behaviour.

## 3.2   Development during childhood

We conceptualise development as a continuous, multidirectional process of change in human functioning that tends to take place to help individuals adapt progressively to contextual demands surrounding them [57, 58, 59, 60]. Within such a multidirectional and contextual conceptualisation of development, children may be better understood as whole human beings with their own capacities rather than as 'incomplete adults' – an alternative conception derived from unidirectional (stage-based) and universal conceptualisations of development [61, 62].

Children's development is supported and constrained by biological and social conditions [63, 64], just like it happens in adulthood – see [65, 66, 67] for biological conditions of adult development; and [68, 63, 69] for social conditions of adult development. Childhood is special, however, in that it is the first period of us human beings in the world. These first years of human life are strongly constrained by the biological maturation of certain parts of the brain [70, 71] and the body [72, 73], whilst also characterised by an enormous quantity of neuronal activity making it the stage of most dramatic developmental changes within the lifespan [74]. Childhood is also a period filled by a strong motivation to learn the socially constructed ways of participating in the world, with children being motivated to learn and develop in ways that allow them to function more autonomously within their specific social contexts [75, 76]. Technology is a central part of human everyday context. It has a strong effect on our actions as it can impose on us its ways of operating [77], which represent social practices, values and systems developed through history by us [77, 38]. Therefore, a conceptualisation of development as multidirectional and contextual is also productive for an enquiry into the affordances of current phygital play technologies on developmentally relevant child behaviour.

## 3.3   Goal-oriented and tool-mediated action analytical framework

We use a *goal-oriented and tool-mediated action* theoretical framing to explore the relationship between play, technology and developmentally relevant child activity.

Such a perspective originates early in the writings of Vygotsky [39], who considered play to generate a zone of proximal development (ZDP) for children. That is, a space that promotes children's *engagement* in activities, thinking processes, and understandings which they would not be likely to engage otherwise, and which could have a positive developmental effect for them [48]. It is, therefore, a theoretical position supporting the idea that play leads to children's development because of how it makes them practice (improving or reinforcing) their own developmentally relevant capacities [78].

Building on Vygotsky's theory is the theory of *perceptual affordances* by James Gibson [79]. This theory is instrumental to understand people's behaviours when using technology. Within developmental psychology, the theory has been translated into the conceptualisation of *action possibilities* [80], or the idea that certain tools or contexts can make some actions more likely than others, given certain developmental capacities [81]. Objects are therefore associated with perceived possible actions [32].

These theories relate to socio-cultural perspectives about technology and human development. They highlight the importance of the social realm in giving technology its developmental function. A very relevant referent within this type of perspective is the work by Aleksei Leont'ev. His work started from the assumption that psychological processes originate from *meaningful object-related activity* [82]. His experiments led him to conclude that the appropriation of social meanings originate from *activity with objects* (objects conceptualised as either the stimuli, tool or aim of human activity) in association with social interactions [37]. This is similar to the Vygotskian perspective indicating that objects do not have any role for the mind if not through their given meaning [83, 37]. It also relates to neo-Vygotskian authors who suggest that the way in which meanings of symbolic tools are appropriated as psychological tools is strongly influenced by the goal or purpose of their use [84].

Indeed, some theorists of the psychology of technology consider the effects *of* and achieved *throughout* technology over the human mind to be very much dependent on the purpose for which we use technology [44]. They consider technological artefacts to be tools modified through history in order to be used in goal-oriented (i.e. intentional) human actions [38]. They also consider *goal-oriented and tool-mediated actions* to be keys to understand the way in which the human mind develops [85, 86]. To understand the purpose given to the use of tools, one needs to consider the nature of the activity setting (such as play). Leont'ev indicates that activity settings can guide actions and determine their functional significance [86].

We also found theories about how technology can force certain types of behaviours due to its operations as well as its physical features. Within play, forcing operations can be widely found in the form of game mechanics [87]. "Game mechanic is an activity structure that consists of rules and the actions afforded to players by those rules" (p.223) [88]. In keeping with this understanding of game mechanics, Arnab et al. [89] and Ke [88] indicate that for a game mechanic to be conducive towards learning, it needs to make players' actions consistent with learning processes. Different empirical reviews have also concluded this by finding that people develop the specific skills repeatedly practiced [90] or simulated [91, 92] within games. Therefore, play mechanics can lead to development when the actions they repeatedly force or constrain are developmentally relevant.

Finally, in relation to physical features of technologies, Donald Norman [93] suggests that control surface interfaces can constrain or even force user behaviour. These effects are achieved at least through two features: the intrinsic properties of their surface representations (e.g., round and elongated shapes permit different uses), and the forcing functions (operations) of the artefact (e.g., car ignition switches require a matching key to work). Such a body of literature also acknowledges that material artifacts are generally charged with cultural or arteficial meanings [38, 94]. Therefore, the use of material artefacts relies on physical constraints and shared conventions that guide their use by people [95, 93].

Based on the above *goal-oriented and tool-mediated action* theoretical framing and other theoretical insights about features of technologies affording user actions, we identify three main facets of play technology (tool) engagement that could explain child behaviour: i) children's *goals* of play technology use (purpose); ii) the digital *mechanics of play technologies* (including interactive rules and users' actions required by those rules); and iii) the *physical characteristics of play technology objects* (material design) constraining users' actions. Overall, we arrive to a framework similar to the Activity Theory Model for Serious Games [96], which identifies *goals*, *tools* and *actions* as main components to analyse the pedagogical uses of serious games. From the theoretical perspective presented above we think we can say the same about developmental effects of 'non-serious' play technologies.

# 4 Methods

## 4.1 Systematic review

We applied a systematic review methodology in order to conduct the review. Systematic reviews are instrumental in making sense of a whole body of literature in order to answer specific research questions [97]. Systematic reviews follow standard procedures during the review process which allow for their replication and updating [98]. Also, they are the only existing evidence-synthesis mechanism lowering the biases of review results by requiring reviewers to assess the bias and quality of reports of individual studies in a transparent way [99]. Moreover, we consider this to be an ideal method because it includes strategies to synthesise studies from a mix of quantitative and qualitative methodologies [100, 101], just as the mix we found in our target field.

We carried out our literature review following the PRISMA guidelines [102] and the general guidance offered by Gaugh, Oliver, and Thomas [98] for systematic reviews. We also considered other complementary guidelines and recommendations when conducting some specific steps of the review, such as the assessment of evidence quality [103], data extraction [104], and narrative synthesis [101, 105, 106].

### 4.1.1 Selection criteria

Papers were included only if considered to help answering RQ.1 fully or partially. We focused on RQ.1 (*What type of social, emotional, cognitive or physical developmentally relevant actions do phygital play technologies afford for children?*) for selection purposes because RQ.2 could also be answered from any study providing information for RQ.1.

In terms of what we considered to be developmentally relevant actions, we focus this review on behaviours that relate to transferable psychological skills (cognitive, emotional and social) which assist adaptation across situations, tasks, or content domains [107]. We also included physical activity as an ubiquitous developmentally relevant behaviour due to its high relevance to the development of cognitive and emotional skills [108, 109]. It is important to note, however, that a contextualist understanding of development, such as the one adopted within this review, might potentially lead to considering any behaviour valued (and therefore studied or promoted) by a particular community (e.g., of researchers, practitioners) to be developmentally relevant. Notwithstanding, for strategic reasons and in order to make the review more relevant for a wider audience and contexts, we focused on transferable skills and physical activity. Consequently, we excluded from this review other types of behaviours, such behaviours indicative of domain- or task-specific skills (e.g., painting, drawing, singing, playing a particular sport) or more academic skills (e.g., mathematics, reading, writing), and attitudinal behaviours (e.g., eating behaviours, sustainable behaviours, motivational behaviours) that children might have engaged in during phygital play.

We only included articles if they met all the following criteria:

- Included measures or observations indicative of transferable developmentally relevant skills and physical activity
- Studied analogue behaviour in digitally enhanced contexts (hence excluding, for example, video-gaming behaviours carried out exclusively within virtual environments)
- Studied play as an activity explicitly carried out for its own sake (rather than an extrinsic goal such as learning)
- Provided results in relation to typically developing 0 to 12-year-olds
- Reported on empirical data
- Were published in English
- Were peer reviewed

## 4.1.2  Search strategy

PsycINFO (all years, until May 15, 2018), ERIC (all years, until May 15, 2018), and ACM Library (from Jan 2013 to June 13, 2018) were searched as the most comprehensive databases for Psychology, Education and Computer Science, respectively. We did not limit the years of our searches in PsychINFO or ERIC. However, following Barr & Linebarger [4] who indicate the high speed of change of new play technologies, we did focus the search of the Computer Science database (ACM Library) on the 5 years leading to the review search (2013-2018) rather than all years.

### 4.1.2.1  Search terms

We framed the exploration of databases' controlled language or thesaurus around three conceptual fields: 'Play', 'Human-computer interactions', and 'Children's development'. In order to achieve a high ratio of relevant hits, we scanned the relevance of the first 40 titles of a series of search simulations and identified syntaxes that produced the highest number of relevant entries.

Unlike PsycINFO and ERIC databases, at the time of the search, the ACM Library search engine (updated and changed in December 2019, after our search) only allowed for the use of partial Boolean logic. That is, it allowed for the logical combination of AND, OR, and NOT commands within but not between search phrases (or lines) making up a full cohesive search syntax. In order to overcome this issue, different end-terms were applied on top of the same root-terms combinations in different single searches within the ACM Digital Library. The searches within ACM Library also reflected the fact that its thesaurus (The ACM Computing Classification System 2012) had much more technology specific terms:

ERIC (via EBSCO): Play AND ("Artificial intelligence" OR "Man Machine Systems" OR "Human Computer Interaction" OR "Handheld devices") AND Child*

PsycINFO (via Ovid): (Play OR "Childhood play behavior" OR "Recreation" OR "Childrens recreational games") AND ("Human Computer Interaction" OR "Human Machine Systems" OR "Mobile devices") AND Child*

ACM Library (via ACM): Search 1 = (Play +Child* +"Human-centered computing"); Search 2 = (Play +Child* +"Tablet computers"); Search 3 = (Play +Child* +"Mobile devices"); Search 4 = (Play +Child* +Smartphones); Search 5 = (Play +Child* +Psychology)

Additionally, we conducted a second search stage on targeted academic journals and conferences considered to publish on the links between technology and human behaviour and which latest's issues or proceedings were not already indexed by our target databases before June 13th 2018. We applied a merge of our psycINFO and ERIC search terms to conduct the search in:

i) *International Journal of Child-Computer Interaction* (from May 2017)
ii) *International Journal of Artificial Intelligence in Education* (from September 2017);
iii) *Personal and Ubiquitous Computing* (from April 2018);
iv) *International Journal of Human Computer Studies* (from April 2018);
v) *Computers in Human Behaviour* (from May 2018);
vi) *International Journal of Human Computer Interaction* (from May 2018);
vii) *Proceedings of the Artificial Intelligence in Education Conference Series* (from 2013), and;
viii) *Proceedings of the European Conference on Technology Enhanced Learning* (from 2013).

We noticed that all conference proceedings indexed in ACM Library were available immediately before or after each conference took place, and were subsequently up-to-date. Therefore, we did not conduct a specific search of relevant conferences indexed in ACM Library such as the Interaction Design and Children conference or CHI Play conference. Furthermore, most of the results yielded by the targeted search (74 out 97 entries) came from conference proceedings not indexed in the ACM Library (sources vii and viii listed above) rather than journals. The targeted search of other relevant academic journals usually produced only between 1 and 3 new entries per journal. Given the low cost–benefit of this approach among journals, we limited its use to only some of the most relevant academic journals in the field.

The total amount of hits per search were 72 (ERIC), 89 (PsycINFO), 361 (ACM 1), 6 (ACM 2), 31 (ACM 3), 30 (ACM 4), 26 (ACM 5), and 97 (targeted journals and conferences). This is equivalent to 712 entries, 635 after duplicate deletion.

### 4.1.3 Study selection

Two reviewers (psychology specialists, first two authors of the present study) screened the titles and abstracts of the resulting 635 papers to determine their eligibility in terms of how they provided information to answer RQ.1. First, a randomly selected 11% (n=70) of the entries underwent a parallel selection process between the two reviewers. Inter-rater agreement was acceptable between reviewers (Cohen's Kappa = 0.62). Following this process, titles and abstracts of the remaining entries from the initial 635 were analysed and recommended for inclusion or exclusion by either reviewer. Reviewers stated applicable exclusion criteria. For a full data base with reasons for exclusions for all entries go to: http://dx.doi.org/10.17632/f5y2vtrnzh.2 [110]. A total of 123 papers were recommended for inclusion (i.e. pre-selected). To strengthen consistency, all pre-selected papers were cross-checked between the two reviewers applying the list of inclusion/exclusion criteria screening the full main texts. Articles which did not meet all the inclusion criteria were excluded from the review. Further, among the 123 pre-selected entries, articles that did not specify sample size and/or age range were also excluded from the review. This exclusion ensured the results of the review could be contextualised in relation to broad participants' characteristics and the size of studies. Also, articles which were too short to get a good sense of their study (i.e. extended abstracts) or quantitative studies including very small sample sizes (10 or less) were excluded from these pre-selected entries to enhance the rigour of the conclusions from the review. Table 1 shows a summary of the reasons for exclusion of all 635 entries (entries could be excluded for more than one reason).

| Reason for exclusion | n |
|---|---:|
| Entries not related to play for its own sake | 313 |
| Entries not connected to outcomes of interest | 309 |
| Entries not analysing analogue behaviour in digitally enhanced contexts | 214 |
| Entries not studying typically developing 0-12 year olds | 161 |
| Entries that were not empirical studies | 96 |
| Entries not specifying sample size and/or participants' age* | 24 |
| Entries that were extended abstracts or did not provide enough information about study* | 16 |
| Entries that were quantitative studies with 10 or less cases* | 5 |
| Entries that were not peer reviewed* | 3 |
| Entries that were not published in English | 0 |

*Criterion coded over full text of pre-selected entries only

*Table 1 Summary of reasons for exclusion*

Figure 1 shows a diagram with the flow of reasons-and-quantities of exclusions from the initial 710 entries identified to the 31 academic articles included for assessment and the 17 included for final review.

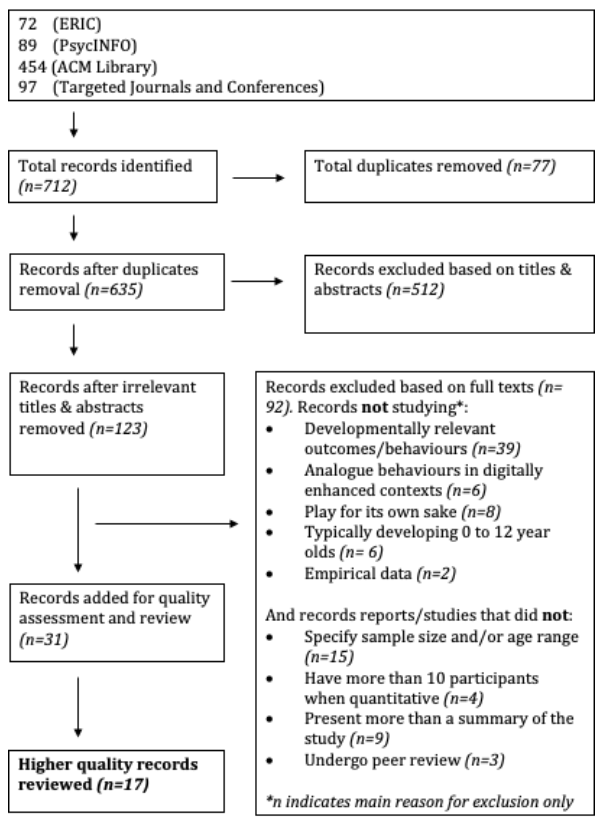

*Figure 1 Papers Review Process Flow Chart*

[figure in black and white]

### 4.1.4   Data extraction

As suggested by Mathes, Klaßen and Pieper [104], in order to reduce bias, extraction of relevant information was conducted in parallel by two independent analysts using the same predefined extraction form. This included key fields to help answer the reviews' research questions, such as research objectives, type of technology, tech functions available to children, and main findings of the study (see full list on appendix A). Other more descriptive information, such as the sample's characteristics, setting of the study, and methods was extracted by one analyst but checked for consistency by a second (see full list on appendix B).

From our data extraction, it was clear that the studies included for the final review were carried out almost exclusively in Western contexts of the world. Out of the 17 studies, most were conducted in the USA (8), or other contexts with strong features of Anglo culture, such as Canada (2), Australia (2) and England (1). The rest of the studies were carried out in European contexts such as The Netherlands (2), Spain (1) and Switzerland (1). Only one study was carried out outside the Western world, in Pakistan (which was also a comparative study with children from The Netherlands). Only a handful of the studies provided more information about the socio-cultural features or demographics of participating children. With the exception of two studies, if gathered, the information was never used for analysis and, therefore, could not be considered for the analytical synthesis of this review. The only type of socio-cultural or demographic information reported in a systematic way for most studies, but again not used for analysis, was the gender of participating children. In the 15 studies that did report the distribution of their sample by sex, participating children tended to be balanced between girls (n=184) and boys (n=187). Further detail about the country and sex of children participating in the studies reviewed can be found in Table 2.

| Study | Country of participants | Sex of participants |
|---|---|---|
| Andrist et al., 2013 | USA | 17 girls / 16 boys |
| Bai et al., 2015 | England | 8 girls / 6 boys |
| Cibrian et al. 2016 | USA | 9 girls / 13 boys |
| Cohen et al., 2014 | Canada (most likely)* | 2 girls / 7 boys |
| Garde et al., 2016 | Canada | 16 girls / 36 boys |
| Hiniker et al., 2017 | USA | 10 girls / 4 boys |
| Hiniker et al., 2018 | USA | 11 girls / 4 boys |
| Hunter et al. 2014 | USA (most likely)* | 7 girls / 5 boys |
| Lawrence, 2018 | USA | 9 girls / 11 boys |
| Malinverni et al., 2018 | Spain | 22 girls / 14 boys |
| Martin-Niedecken, 2018 | Switzerland | 16 girls / 16 boys |
| McKenzie et al., 2014 | Australia | 5 girls / 9 boys |
| Saksono et al., 2015 | USA | 10 girls / 4 boys |
| Shahid 2018 | The Netherlands | Mixed gender - n unspecified |
| Shahid et al. 2014 | Pakistan & The Netherlands | Mixed gender - n unspecified |
| Shen et al., 2018 | USA | 34 girls / 30 boys |
| Straker et al., 2009 | Australia | 8 girls / 12 boys |
| TOTAL | | 184 girls / 187 boys |

*Indicates the country of University affiliation of first author and most co-authors when no other information was provided about location of study

*Table 2 Country of study and sex of study participants*

## 4.1.5 Assessment of quality of study reports

In order to assess the quality of the evidence, two analysts applied 10 and 11 quality assessment criteria for quantitative and qualitative studies respectively. An adapted version of the *Checklist for Randomised and Non-randomised Studies* [111] was applied for assessment of quantitative evidence. The *Checklist for Qualitative Research Quality Appraisal* [112] was applied for assessment of qualitative evidence (see versions applied in Appendix C). Mixed methods studies were assessed using both assessment frameworks. All papers were assessed by two reviewers and all assessment differences were discussed and agreed, reaching a 100% consensus. A detail of the quality assessment per paper is offered in Table 3 and Table 4 for quantitative and qualitative evidence, respectively. Mixed methods studies (*) appear in both tables.

| Article | Reporting | | | Internal validity | | | | | | | External validity | Overall quality | | |
|---|---|---|---|---|---|---|---|---|---|---|---|---|---|---|
| | Aim | Clear D.V. | Findings | Blinding | Measures | Comparability | Randomisation | Stats | Confound control | Control of adult effect | Representativity | Score | Max. applicable | Quality Index |
| **Higher quality studies included for synthesis** | | | | | | | | | | | | | | |
| Bai et al. [113] | YES | YES | YES | YES | YES | YES | NA | YES | YES | | | 8 | 10 | 8.00 |
| Shahid [114] | YES | YES | YES | | YES | YES | | YES | YES | YES | | 8 | 11 | 7.27 |
| Andrist et al. [115] | YES | YES | YES | | YES | | NA | YES | YES | YES | | 7 | 10 | 7.00 |
| McKenzie et al. [116] | YES | YES | YES | | YES | NA | NA | YES | | YES | | 6 | 9 | 6.67 |
| Shahid et al. [117] | YES | YES | YES | YES | YES | | | YES | YES | | | 7 | 11 | 6.36 |
| Shen et al. [118] | YES | YES | YES | | YES | | YES | YES | | YES | | 7 | 11 | 6.36 |
| Garde et al. [119] | YES | YES | YES | YES | | | NA | YES | | YES | | 6 | 10 | 6.00 |
| Hunter et al. [120] * | YES | YES | YES | | YES | YES | NA | | | YES | | 6 | 10 | 6.00 |
| Straker et al. [121] | YES | YES | YES | | YES | | NA | YES | | YES | | 6 | 10 | 6.00 |
| % of HQ studies meeting criterion (if applicable) | 100% | 100% | 100% | 33% | 89% | 38% | 33% | 89% | 44% | 78% | 0% | | | Avg:6.63 |
| **Lower quality studies excluded from synthesis** | | | | | | | | | | | | | | |
| Chaspari et al., 2015 [128] | YES | YES | YES | | | NA | NA | YES | | YES | | 5 | 9 | 5.56 |
| Sugimoto, 2011* [129] | YES | YES | YES | | YES | NA | NA | | | YES | | 5 | 9 | 5.56 |
| Castañer et al., 2016 [122] | YES | YES | YES | | YES | YES | | | YES | | | 6 | 11 | 5.45 |
| Malinverni et al., 2018* [123] | YES | YES | | | YES | YES | YES | | | | | 6 | 11 | 5.45 |
| Tewari & Canny, 2014 [124] | YES | | YES | | | YES | YES | YES | YES | | | 6 | 11 | 5.45 |
| Kerepesi et al., 2006 [127] | YES | YES | | | YES | YES | | YES | | YES | | 6 | 11 | 5.45 |
| Boccanfuso et al., 2016 [125] | YES | YES | YES | | | | NA | | YES | YES | | 5 | 10 | 5.00 |
| Cohen et al., 2014* [126] | YES | YES | YES | | YES | YES | NA | | | | | 5 | 10 | 5.00 |
| Martin-Niedecken, 2018* [130] | YES | | | | YES | YES | | | | YES | | 5 | 10 | 5.00 |
| Jeong et al., 2017 [131] | YES | | YES | | | | YES | | | | | 3 | 11 | 2.73 |
| % of LQ studies meeting criterion (if applicable) | 80% | 50% | 60% | 0% | 50% | 75% | 60% | 30% | 30% | 30% | 0% | | | Avg:5.07 |
| Total of articles meeting criterion (if applicable) | 100% | 84% | 89% | 16% | 74% | 56% | 50% | 63% | 37% | 63% | 0% | | | Avg:5.81 |

Legend: Aims = Clear purpose; Clear D.V. = Clear outcome variables; Findings = Clear findings; Representativity = Representative sample; Blinding = Blinding researchers measuring outcomes to purpose or group allocation; Measures = Appropriate measures; Comparability = Comparability of conditions; Randomisation = Participant randomisation; Stats = Appropriateness of statistical tests; Confound control = Control of confounding variables; Control of adult effect = No risk of outcomes being confounded with adult participation in play situation.
YES= Study clearly meets criterion; NA= Criterion does not apply to study; UTD= Unable to determine if study meets criterion due to lack of information; * Mixed methods articles assessed using quantitative and qualitative criteria.

*Table 3 Quality Assessment of Quantitative Evidence*

| Article | Reporting | | | Methods | | | | | Reflexivity and ethics | | Total |
|---|---|---|---|---|---|---|---|---|---|---|---|
| | Aims | Value | Findings | Qual. | Design | Recruit. | Data collection | Data analysis | Rel. with participants | Ethics | Score ( /10) |
| Higher quality studies included for synthesis | | | | | | | | | | | |
| Hiniker et al. [132] | YES | YES | YES | YES | YES | YES | YES | YES | YES | YES | 10 |
| Hiniker et al. [133] | YES | YES | YES | YES | YES | YES | YES | YES | | YES | 9 |
| Lawrence [134] | YES | YES | YES | YES | | | YES | YES | YES | YES | 8 |
| Martin-Niedecken [130] * | YES | YES | YES | YES | YES | YES | YES | | | YES | 8 |
| Malinverni et al. [123] * | YES | YES | YES | YES | | | YES | YES | | YES | 7 |
| Saksono et al. [135] | YES | YES | | YES | | YES | YES | YES | | YES | 7 |
| Cibrian et al. [136] | YES | | YES | YES | | | YES | YES | | YES | 6 |
| Cohen et al. [126] * | YES | YES | YES | YES | | | YES | YES | | | 6 |
| % of HQ studies meeting criterion (if applicable) | 100% | 88% | 88% | 100% | 38% | 50% | 100% | 88% | 25% | 88% | Avg: 7.62 |
| Lower quality studies excluded from synthesis | | | | | | | | | | | |
| Hooft van Huysduynen et al. [137] | YES | YES | | YES | | | YES | | YES | | 5 |
| Hunter et al. [120] * | YES | YES | YES | YES | | YES | | | | | 5 |
| Hoare et al. [138] | YES | | | YES | | | | | | YES | 3 |
| Rogers & Muller [139] | YES | YES | | YES | | | | | | | 3 |
| Sugimoto, 2011 [129] * | YES | YES | | YES | | | | | | | 3 |
| Goh et al. [140] | YES | | | YES | | | | | | | 2 |
| Kozima et al. [141] | YES | | | YES | | | | | | | 2 |
| Jeong et al. [142] | | | | | | | | | | | 0 |
| % of LQ studies meeting criterion (if applicable) | 88% | 50% | 13% | 88% | 0% | 13% | 13% | 0% | 13% | 13% | Avg: 2.88 |
| Total of articles meeting criterion (if applicable) | 94% | 69% | 50% | 94% | 19% | 31% | 56% | 44% | 19% | 50% | Avg: 5.25 |

*Legend: Aims = Clear communication of aims; Value = Presentation or discussion of study's value; Findings = Clear communication of findings; Qual. = Fit of qualitative methods; Design = Appropriate research design; Recruit. = Appropriate recruitment strategy; Data collection = Appropriate data collection methods; Data analysis = Rigorous data analysis; Rel. with participants = Considers researcher-participant relationship; Ethics = Considers research ethics.*

YES= Study clearly meets criterion; * Mixed methods articles assessed using quantitative and qualitative criteria.

*Table 4 Quality Assessment of Qualitative Evidence*

As can be seen from Table 3 and Table 4, available evidence varied in terms of quality, regardless of its methodological approach. Articles averaged 5.25 and 5.81 in a quality index out of 10 points in both qualitative and quantitative studies, respectively. We focused our synthesis analysis only on higher quality papers (meeting 60%+ of quality criteria) in order to make sure the quality of reported evidence did not bias or discredit the results of the review. Higher quality qualitative and quantitative papers averaged 7.62 and 6.63 out of 10 in the quality index, respectively, making qualitative evidence the strongest.

### 4.1.6   Analytical method for evidence synthesis

Due to the large heterogeneity of evidence presented by the articles reviewed, and following the guidance of the Cochrane Handbook (Version 5.1, section 20.3.2.4) [101], we applied a narrative synthesis for evidence synthesis. Narrative synthesis is considered to be particularly appropriate when synthesising studies that are either insufficiently similar (in terms of participants, interventions, outcomes) to allow for a specialist synthesis, or follow a range of research design that produce quantitative and qualitative findings requiring synthesis [105]. Both these types of diversities were found among our selected studies. In order to conduct the synthesis, we partially followed the guidance by Popay et al. [105] to make the process more systematic. This included two stages: 1) establishing a theoretical framework (*goal-oriented and tool-mediated action* framework), and 2) analysing and exploring relationships in the data based on the theoretical framework. The analysis required for stage 2 was carried out by the first and second authors (reviewers) first in parallel and then jointly.

# 5   Results

Following our research questions, we focused the analysis on trying to understand how play technology afforded children's development. Specifically, following the theoretical framework presented above, we aimed to find patterns in terms of the ways in which play technologies (*tools*) generated developmentally beneficial *goal-oriented actions* for children. To carry out this analysis we took whole studies as a point of reference. We did not limit the analysis to the main research objectives declared by the original researchers. Instead, given the novelty of the field, we considered any insights and results reported by original researchers to make our own overall analysis across studies.

To make our analysis more transparent, throughout the review we explicitly differentiate between insights/results reported by the authors of studies reviewed (e.g., "the authors/the results found/suggest/indicate") and our own inferences or insights (e.g., "we consider/think/infer") as reviewers. Descriptions of the main research objectives, play situations and technologies, research methods, and findings declared by authors themselves can be seen in Appendix D.

## 5.1   Analytical narrative synthesis

Our analysis suggests that there are four different types of goal-oriented and tool-mediated developmental actions that play technologies promoted among children. We considered these to be developmentally beneficial in the sense that they allowed children to practice transferable skills [107] valued (studied) within developmental and educational psychology or promoted physical activity, a type of ubiquitous behaviour known to promote cognitive and emotional development [79, 80]:

1) Facilitating child self-monitoring;

2) Promoting collaboration;

3) Inviting children's decision making;

4) Forcing problem solving and physical activity.

Same play technologies were found to afford more than one of these developmentally relevant behaviours.

As will become clear in the next sections, these developmentally relevant behavioural affordances emerged at the interplay between a variety of digital play mechanics, physical characteristics of technology, social expectations and the play goals pursued by children.

### 5.1.1   Facilitation of children's self-monitoring (6 studies)

Self-monitoring is a metacognitive, executive function involving on-line reflection on, and evaluation of, cognitive processes. It forms a part of self-regulation, which in-turn is defined as the ability to plan, monitor and adapt one's own behaviour [143]. In conjunction with student goals [144], self-monitoring generates a feedback cycle with which to manage and regulate cognitive processes. As such, proficient self-monitoring is assumed to contribute to improved cognitive performance [145]. The process allows the child to monitor and control their actions personally and independently, and thereby acts as a self-improvement tool [146] for any type of developmental and learning aim [147].

Within this review, facilitation of children's self-monitoring refers to how play technologies promoted children's awareness of their own performance towards meeting play goals and rules, either contingently or asynchronously. Studies supporting this finding included those by Cibrian, Weibel, and Tentori  [136], Garde et al. [119], Hiniker, Lee, Sobel, and Choe  [133], Saksono et al.  [135], Shen, Slovak, and Jung  [118] and McKenzie, Bangay, Barnett, Ridgers, and Salmon  [116]. Table 5 shows a summary of these studies, including only results relevant to self-monitoring.

| Author & year | Objective (To study) | Play Goals | Key findings (relevant to review results) |
|---|---|---|---|
| Cibrian et al. 2016 | Fabric-based interactive surfaces impact on children's development | Play with the interactive fabric during free-play in their classrooms. Children use movement and touch to erease nebulas, reveal underlying space elements and play music | • Children improved in their motor skills<br>• Children improved their sustained attention<br>• Children adjusted the pressure they put on the fabric in order to keep their balance whilst playing with the screen* |
| Garde et al., 2016 | Exergame impact on children's school-based physical activity | Children accumulated points for a mobile videogame by means of physical activity outside the videogame environment | • Physical activity increased in equal degrees among boys and girls<br>• The exergame helped children monitor the amount of physical activity collected to fuel their videogame, hence incetivicing more physical activity to fuel play* |
| Hiniker et al., 2017 | Technology impact on children's self-regulation and parents' support | Children planned and followed their own tablet play plan with supervision from parents | • Children showed clear intentionality when planning their games<br>• Children self-regulated successfully, keeping to their plans in 93% of planned transitions even though they were not forced to make the transition<br>• Both parents and children treated the app as a third party authority |
| McKenzie et al., 2014 | Mobile-phone game impact on children's outdoor physical activity | Children played a treasure hunt game using their mobiles. Each player received a a map with a set of predefined clues and movement activities. Players interacted with gameplay by hunting for treasure locations (QR codes) with a set of predefined clues and movement activities challenges | • Most children found the level of difficulty to be between "just right" and "too hard"<br>• Information of player's relative position to other players made children go faster or take alternative routes between clues to beat competitors |
| Saksono et al., 2015 | Collaborative exergame impact on children's and parent's physical activity | Children and caregivers engaged in physical activity to earn time and points of a videogame in order to complete space missions | • It increased awareness of opportunities for excercise and motivated those already aware of physical activity opportunities to follow them up<br>• Caregivers and children collectively assessed their physical activity reached during the game |
| Shen et al., 2018 | Social robot impact on children's interpersonal conflict and resolution skills | Children played the games of choice from: Lego Duplo, magnetic tiles, toy house, remote control car, and making a birthday card across two conditions. First facilitated and directed by a robot, second also mediated by the robot | • There were no differences in socialness, constructiveness or amount of conflicts across conditions<br>• When the social robot mediated conflicts, children were 4 times more likely to resolve conflicts in positive ways<br>• The pausing of play enforced by the social robot following a social conflict helped children resolve their conflicts |

\* Information inferred by reviewers from the report of each study

*Table 5 Summary of studies relevant to self-monitoring*

### 5.1.1.1 Providing information about meeting play rules (3 studies)

The main feature facilitating children's self-monitoring was providing information about own performance. In particular, technologies provided children with *contingent feedback about own rule-related behavioural performance.* This contingent feedback was observed in the studies by Hiniker et al. [133], Shen et al. [118], and Cibrian et al. [136].

Hiniker et al. [133] reported on 11 four to six year-old children's use of the Plan&Play tablet app. The app was designed to allow children to select tablet games to play, order them, and allocate different lengths of time to play them (i.e., setting own rules of play). The app provided contingent information about children's progress-status in relation to their own plans using widgets. Results show that when children used Plan&Play they followed the order and time allocation they planned 93% of the time. Given that the app did not force children to make game transitions, we consider that the provision of contingent information about own play status was sufficient to facilitate their self-monitoring to meet their own plans.

Shen, Slovak, & Jung [118] also showed how contingent feedback engaged children in self-monitoring to meet play rules. The study looked at how 64 three to six year-old children reacted to Keepon robot when it moderated children's social conflicts (i.e. play harmonically with others) within play. The authors found that the key factor helping children being able to resolve social conflicts was that Keepon paused children's play when sensing conflict, hence informing children of their rule-related behaviour.

In addition, Cibrian et al.'s study [136] looked at 22 two to three year-old children's engagement with Bendable Sound. The Bendable was a large, standing, bending "touch" screen (a flexible fabric connected to a Kinect recognising children's touch movements). Researchers observed that when children's touch was too strong, the Bendable would not generate the desired play effect for children, making them lose their balance against the soft fabric. This allowed children to adjust to the appropriate pressure expected (i.e. material rule) when operating the device.

### 5.1.1.2 *Providing information about progress towards goals (3 studies)*

Self-monitoring was also facilitated by *providing information about own progress towards play goals.* In Saksono et al.'s and Garde et al.'s studies, technology invited 14 eight year-old children [135] and 28 nine to thirteen year-olds [119] to self-monitor to achieve intermediate play goals such as gathering fuel to play video-games by engaging in physical activity. Similarly, in the study by McKenzie et al. [116], the information about the player's own relative position to other players made 14 five to twelve year-old children go faster or take routes between clues to beat competitors.

## 5.1.2   Promoting collaboration (12 studies)

Collaboration, in developmental psychology terms, can be understood as coordinated and synchronous activity with another in an attempt to develop and maintain a shared understanding [148] or to achieve a common goal [149, 150].  The value of collaboration to cognitive development was described by both Piaget [151] and Vygotsky [152]. Piaget suggested cognitive development occurred when the child, through dialogue and discussion, attempts to resolve a discrepancy in knowledge between the self and others.  Vygotsky suggested that development tends to occur when a child, through meaningful social interactions, arrives at a shared understanding with a more knowledgeable peer or adult. More recent work indicates that collaboration could be directed to help develop other specific skills, such as reading, reasoning, and learning [153, 154, 155] as well as help to achieve any type of goal that one person could not achieve alone [150]. Knowing how to collaborate, however, is not a given and there are many skills involved in being able to collaborate, such as knowing how to manage joint group attention, establish common knowledge for a task, negotiate with others, take turns, and regulate conflicts [150]. Evidence suggests that children of all ages obtain benefits from collaboration [156]. Also, in keeping with our findings, previous work has indicated that collaboration can be facilitated through task design [150].

Within this review, promoting collaboration refers to how play technologies invited children to co-construct, negotiate or coordinate joint play. Technologies delivered this by either slowing the pace or pausing play in social situations; making play interdependent between players; or facilitating joint accessibility of playmates to play objects. We observed the creation of these collaborative spaces in the studies by Cohen, Dillman, MacLeod, Hunter, and Tang [126], Hiniker, Lee, Kientz, and Radesky [132], Lawrence [134], Shen et al. [118], Malinverni, Valero, Schaper, and Pares [123], Martin-Niedecken [130], Bai, Blackwell, and Coulouris [113], Saksono et al. [135], Cibrian et al. [136], Andrist, Leite, and Lehman [115], Hunter, Maes, Tang, Inkpen, and Hessey [120] and Shahid [114]. Table 6 shows a summary of these studies, including only results relevant to collaboration.

| Author & year | Objective (To study) | Play Goals | Key findings (relevant to review results) |
|---|---|---|---|
| Andrist et al., 2013 | Virtual agent impact on children's turn taking, overlapping speech, enjoyment | Children use an interactive language-based game to change the appearance of a model by calling out the names of items on the game board. Their play is mediated by a virtual character responsible for making the changes and mediating the interaction between children | • The more flexible virtual agent able to apply actively all four types of turn-passing moderation strategies (gesture, gaze, proxemics and speech) was found to lead to more equal participation during play<br>• Children's talk tended to overlap regardless of the type(s) of the moderation strategies use by the virtual agent |
| Bai, Blackwell, & Coulouris, 2015 | Augmented reality system impact on children's joint pretence, emotion expression, divergent thinking | Children use the magic mirror AR to create stories by manipulating physical referents and assign meaning to them (e.g., pirate), props (e.g., bicycle) and scenary textures (e.g., grass) | • The AR system elicited large quantities of pretence play across all children<br>• Children engage in a variety of types of communication to coordinate social pretend play |
| Cibrian et al., 2016 | Fabric-based interactive surfaces impact on children's development | Play with the interactive fabric during free-play in their classrooms. Children use movement and touch to erease nebulas, reveal underlying space elements and play music | • Children preferred to play together rather than alone (especially when creating sounds together) or when imitating each other / taking turns<br>• Children improved communication skills |
| Cohen et al., 2014 | Video conference system impact on children's nature of distance play interactions | Children played open-endedly through video-conference platforms. One platform was akin to Skype and the other merged the images from both interactuants into one screen which both could see. Toys were made available to them such as stuffed animals, masks, books, drawing materials, large legos, etc. | • The video conference system promoted more engagement in organised and cooperative (i.e. collaborative) play, whereas the conventional configuration promoted more parallel and associative play<br>• The video conference system promoted more active/physical play, discussion and action between players, whereas the conventional configuration promoted more show-and-tell and make-believe play<br>• In the conventional setting parents tended to coach children more and they would usually disengage with one another devolving into parallel or associative play or stopped playing |
| Hiniker et al., 2018 | To better understand the ways in which traditional toys and digital Apps enable and inhibit parent-child play. | Children played open-endedly with their favourite table games and non-digital toys | • Dyads were more likely to engage in joint activity when playing with toys than with tablets<br>• Children tended to put toys, but not tablets, in the attentional spaces shared between them and their parents<br>• Parents were more likely to engage children in turn taking and conversation when playing with toys<br>• Tablets made difficult for parents to see what their children were doing, except when games allowed for multi-touch, which made children share the screen with parents to play together<br>• Children tended to ignore parents' questions when playing tablet games<br>• Children were more likely to trail off from conversations when app games showed visual effects or prompted them<br>• When apps allowed for self-paced activity children were more able to manage their own attention<br>• Traditional toys made children focus more on the play experience while also sustaining conversation with parents |
| Hunter et al., 2014 | Cooperative virtual interaction impact on creative play and social engagement | Children played together open-endedly inside digital environments that they select, create, and arrange | • Being together in the same virtual space was found to be the most effective way of making players engaged in shared activities and being creative. It also tended to increase play engagement and the diversity of play types<br>• Participants focused more on each other when playing under a "mirror" mode<br>• Making puppets the only visible objects in some modes of the technology, and allowing for the creation of joint-assets might have helped collaborative play * |
| Lawrence, 2018 | Collaborative play using tablets impact on children's interactions | Children played a selection of 5 different table apps together | • Children struggled for tablet control, especially in closed-ended reward based games and towards the beginning of play<br>• Children cooperated, collaborated and enjoyed playing in the open-ended play app<br>• Fast paced games turned players competitive, but slower paced games and games with pauses in between activities turned players collaborative |

* Information inferred by reviewers from the report of each study

*Table 6 Summary of studies relevant to collaboration*

| Author & year | Objective (To study) | Play Goals | Key findings (relevant to review results) |
|---|---|---|---|
| Malinverni et al., 2018 | Augmented & mixed reality impact on children's understanding, collaboration and use of physical space | Children played using either a shared tablet or portable projector. They found "magical portals" (markers) located on the walls of two different spaces within the school in order to solve a mystery | • Children using the portable projector performed more gestures to express emotions, more verbal exchanges focused on co-constructing knowledge with peers. They also physically arranged themselves in a semi-circle to work together<br>• Children using the shared tablet established clear divisions of roles instead and their physical arrangement was more scattered |
| Martin-Niedecken, 2018 | Motion controllers impact on children's social and bodily interplay | Children played a videogame using two types of body motion controllers: a full body motion controller (FBMC) or a Kinect. The two children controlled the game as if they were one (using split controllers). They operated in the world of a young pirate, searching for buried treasures with a flying ship and overcoming challenges using physical movements. | • Both FBMC and Kinect facilitated some level of collaborative coordinated play<br>• Participants engaged more in interdependent bodily interplay and communication with the FBMC than the Kinect<br>• Participants without sports skills but with gaming skills interacted more confidently and were more cooperative when using FBMC |
| Saksono et al., 2015 | Collaborative exergame impact on children's and parent's physical activity | Children and caregivers engaged in physical activity to earn time and points of a videogame in order to complete space missions | • Caregivers and children collectively assessed their physical activity reached during the game<br>• Most participants preferred to use the game in competitive (comparing not outperforming each other) rather than collaborative ways |
| Shahid, 2018 | Video mediated communication system affect on children's feeling of social presence and socio-emotional response | Children played a computer card game either alone, in collaboration with a friend or with a robot. When winning children collected coins as rewards | • Children's socio-emotional perceptions of co-presence as well as message and affect understanding were higher in the ideal and mutual gaze conditions<br>• Children had least fun, were least expressive and did not feel a social bond in the no-gaze condition |
| Shen et al., 2018 | Social robot impact on children's interpersonal conflict and resolution skills | Children played the games of choice from: Lego Duplo, magnetic tiles, toy house, remote control car, and making a birthday card across two conditions. First facilitated and directed by a robot, second also mediated by the robot | • There were no differences in socialness, constructiveness or amount of conflicts across conditions<br>• When the social robot mediated conflicts, children were 4 times more likely to resolve conflicts in positive ways<br>• The pausing of play enforced by the social robot following a social conflict helped children overcome conflicts and play more collaboratively |

\* Information inferred by reviewers from the report of each study

*Table 6 Summary of studies relevant to collaboration* (continued)

## 5.1.2.1 Slowing and pausing the pace of play (3 studies)

The importance of slowing the pace of play or allowing players to define their own pace of play and adding pauses into play progressions was found in the studies by Lawrence [134], Hiniker et al. [132], and Shen et al. [118]. Lawrence [134] observed how 20 five year-old children played together with different types of tablet video-games. According to the study, tablet applications clearly demarking the beginning and end of different activities through pauses promoted collaboration between children. This was even more evident in Lawrence's [134] observation that when games were fast-paced they tended to turn competitive the very same group of children who had played collaboratively in slower games.

Conversely, Hiniker et al. [132] found that fast interactivity of video-games led 14 children aged four to six to play alone even when encouraged to play with parents. The authors compared parent-child interactions across different types of tablet games and traditional toys. They found that when apps allowed for self-paced activity, children were able to manage their own attention and play collaboratively with their parents rather than being individually absorbed by the interactivity. They also found that children tended to play collaboratively when using traditional (non-digitally enhanced) toys, which also let children self-pace their own play.

Finally, the already described study by Shen et al. [118], where Keepon robot helped 64 three to six year-old children overcome social conflicts by pausing play, also evidenced the importance of pausing to promote collaboration. Authors themselves indicate that pause enforcement was the key factor that had the most effect on promoting children's overcoming social conflicts and engage in more collaborative interactions.

### 5.1.2.2 Making play interdependent (4 studies)

Designing for playmates' interdependency was another way through which play technologies generated collaboration. Interdependency refers to how people need each other to achieve goals, making the outcome of one person directly linked to the outcome of another [157]. Cohen et al. [126] studied how two types of whole-body video-conferencing platforms, namely OneSpace and a Skype-type platform, afforded social play among 9 six to ten year-old children, or between these and their parents. OneSpace was a whole-body video conferencing tool that merged the video feeds of two remote sites into a single shared visual scene. When players positioned themselves at different depth levels within it, the player who was closer to his/her respective depth camera shot was displayed in front of the other player. This effect of co-presence allowed the generation of physical interactive play between playmates, such as chasing one another or playing hula-hoop when their body images merged together. Results showed that the whole-body Skype type platform generated more parallel and associative play. OneSpace, on the other hand, generated more true collaboration between playmates based on physical movements.

In a similar vein, Hunter et al. [120] studied the use of WaaZam, a video mediated communication (VMC) system to support creative play and increase social engagement of geographically separated families. They examined the interactions between children aged 6-12 and adults playing across four different VMC virtual settings: separate windows as conventional videoconferencing; merged 'magic mirror' windows where one person can appear in the others' space; digital play sets where both players are merged into the same fictional environment; and customised digital environments where the players have the option of adding and changing their constructed merged virtual worlds. Results showed that being in the same virtual space increased play engagement and made more shared activities possible. Participants also focused more on each other in mirror mode. Being together in the same virtual space, playing interdependently, emerged as the most important factor to supporting more creative and social activities.

Andrist, Leite, and Lehman [115] developed a fashion game in which a projected virtual agent helped 33 children aged four to ten to take turns in dressing a projected virtual character, using a total of up to 20 turns in a group setting. The study focused on the action of turn-taking. Findings suggest that balanced turn-taking was possible to achieve but only when the virtual agent moderated turns more actively through gesture, gaze, proxemics and speech, rather than when it only applied a sweeping gesture to pass the turn. We consider that when the moderation of turn-taking was actually ensured through more directives, the game became more interdependent not just in the outcome but the process (children clearly had to wait for one another). This interdependency during the process of play, in turn, promoted a more balanced collaboration between players. The play technology studied by Martin-Niedecken [130] also generated interdependency. The play technology required two 32 ten to fourteen year-old children (and young adolescents) to operate the game controls in synchrony as pairs. The author compared such an interdependency having a full body motion controller (FBMC) and a Kinect as controlling devices. When using the FBMC to control the game, participants needed to move and jump hitting large fixed buttons (positioned at low, middle, and high heights near the player's body) with their hands. Alternatively, when using the Kinect, participants needed to move their bodies into different shapes, jump/duck and make specific bodily gestures to control the game. Results showed that whilst both interdependent settings led to collaboration, a FBMC was the most effective.

### 5.1.2.3  *Making objects of social (multiuser) play accessible for joint attention (9 studies)*

Joint accessibility of the objects of social play activity was important to ensure collaboration during play. The work of Martin-Niedecken [130] on FBMC/Kinect controllers, presented just before, provides evidence about the importance of such accessibility to enhance collaboration. Although not concluded by the author himself, we suggest that Martin-Niedecken's [130] study evidences the importance of directing joint attention to very specific objects (pushing buttons) to facilitate players' coordination. We consider that a possible reason why the FBMC was more collaborative than the Kinect is because joint attention was more difficult to achieve when using the Kinect. In particular, when using the Kinect, children were required to turn around their faces between screen and playmate in order to coordinate. This was not the case for the FBMC. The FBMC was placed slightly in front of both children, permitting either child to see both sides of the FBMC at the same time when making use of their full visual range (front and side). This made collaboration easier.

Related research supporting the importance of joint accessibility for collaboration is that of Bai et al. [113]. The authors studied an augmented reality magic mirror tool used by fourteen four to six year-old children to engage in peer pretend play. Two children held wooden or wireframe puppets and shapes which would then be magically transformed into referents of their choice (e.g., pirate, bicycle, grass) on the screen. The authors found that the systems made children engage in a variety of types of communication to coordinate social pretend play. We think this collaborative coordination was facilitated by giving children easy access to objects of joint attention in their hands and the screen.

Hiniker et al. [132] indicate how traditional toys afforded more collaboration than tablets. They observed how fifteen four to six year-old children played with analogue toys and tablet games in the presence of or together with their parents. The study showed that children tended to put toys, but not tablets, in the attentional spaces shared between them and their parents. As a consequence, tablets tended to produce more individual play by children, overseen but not joined by parents. Tablets also afforded less collaboration due to the difficulty of achieving joint attention through their small (one-way facing) screens. The authors observed, however, that the limitations of the tablet screen for collaborative play could be overcome through digital play mechanics. In particular, they found that when tablet games allowed for and invited parallel multi-touch across players, children tended to position the tablet in such a way that helped parents to also see and touch the screen. This led to collaboration.

The effect of tablet screens on collaboration was also found in the study by Malinverni et al. [123]. The study compared the behaviour of 36 nine to eleven year-old children in mixed-reality games delivered by a tablet versus a projector. Children played in groups and used the same game across study conditions. The projector was found to promote collaborative meaning making through co-construction and co-creation among children. This was not observed in the tablet solution. Instead, the tablet lowered the amount of joint engagement and collaboration among children.

Hunter et al. [120] design and explore the use of physical objects for joint attention in shared virtual worlds between 12 six to twelve year-old children and their parents. The WaaZam system 'puppet' mode allows for physical puppets to be the focus of the players by showing only these objects on the screen. When creating virtual scenes, the players also have the possibility to create a library of joint assets. It could be that the availability of these objects for joint attention also helped the collaborative success of WaaZam.

The study by Saksono et al. [135] on the Spaceship Launch exergame (video-game fuelled by physical activity) also succeeded in promoting collaboration between parents and 14 children (8 years of age, average). The game showed both collaborators the contribution that each of them made towards reaching their joint targets. We think that provided players with a target of joint attention. Furthermore, Cibrian et al.'s [136] study on the Bendable Sound found that 22 children aged two to three years preferred to make use of the device in social ways (especially when creating sounds together). This was observed despite the fact that the device and its games were designed for individual play. We think that what made children play together in this fast-paced interactive game was the large size of the Bendable's screen (higher and wider than children themselves). The large screen made the game jointly visible to children, both allowing and inviting them to interact with the screen together through multi-touch.

Shahid [114] investigated how different levels of gaze in a video mediated communication system might affect 108 eight year-old children's feelings of social presence and socio-emotional response. The author observed the children's socio-emotional response whilst playing a collaborative virtual card game remotely across three conditions: ideal mutual gaze, semi-ideal mutual gaze and no gaze at all. The ideal condition allowed players to establish eye contact and mutual gaze, the semi-ideal condition allowed for the ability to look at the other player's face but without eye contact, and finally in the no gaze condition the players could only see the game, not seeing each other. Results show that children reported the feeling of social presence most during the ideal gaze condition and least during the no-gaze condition, and semi-ideal. Also, children showed most non-verbal cues during the ideal gaze condition (i.e. that facilitating joint attention), least during the no-gaze condition, with semi-ideal gaze in between. This is very relevant because non-verbal cues are facilitators of clarity of communication [158] and, therefore, collaboration.

Finally, the study by Andrist, Leite, and Lehman [115] was the only study reviewed that showed inconclusive results in relation to the association between accessibility to joint attention and collaboration. As indicated before, the study focused on promoting turn-taking with different levels of assistance offered by a digital agent in a fashion game projected on a wall. Findings suggest that, despite all conditions promoting joint attention on the agent, they did not all insure collaboration to equal degrees. Having said that, we still think that children could be considered to have collaborated to some extent across conditions. We infer this because their speech overlapped regardless of the type of moderation (condition), indicating that children negotiated their fashion decisions across all conditions. More precisely, we think this study shows that although features designed to encourage joint attention promote children's active exchanges of ideas, they do not necessarily ensure socially balanced collaboration.

### 5.1.3 Promoting children's decision making (5 studies)

Decision making is a complex skill that is influenced by multiple factors, in particular the ability to reason about a decision and control emotional responses elicited by a decision problem [159]. Good decision makers tend to be more successful in their adaptation to the environment [160], and this capacity tends to improve between childhood and adolescence [161]. Development of decision making effectiveness occurs by making decisions, gathering information about what affects decision making, and through reflection [162].

Within this review, promoting children's decision making refers to how play technologies require children to make choices in order to play or progress towards games' end goals. These choices also made children engage in developmentally beneficial thinking and actions. This was observed in the studies by Bai et al. [113], Hiniker et al. [133], McKenzie et al. [116] Saksono et al. [135] and Hunter et al. [120]. Table 7 shows a summary of these studies, including only results relevant to promoting children's decision making.

| Author & year | Objective (To study) | Play Goals | Key findings (relevant to review results) |
|---|---|---|---|
| Bai, Blackwell, & Coulouris, 2015 | Augmented reality system impact on children's joint pretence, emotion expression, divergent thinking | Children use the magic mirror AR to create stories by manipulating physical referents and assign meaning to them (e.g., pirate), props (e.g., bicycle) and scenary textures (e.g., grass) | • Children made explicit and deliverate decisions about their play when presented with options about the characters and play background - they would also change the emotion of their characters if given the choice<br>• Children were found to make more verbal communications of their transformations to playmates when choosing from open-ended representations (e.g., black) rather than from more definite-meaning representations (e.g., pirate)<br>• Children generated more imaginary representations with familiar than with less familiar scenary textures, but when working with less familiar materials they tended to be more novel in their imagination |
| Hiniker et al., 2017 | Technology impact on children's self-regulation and parents' support | Children planned and followed their own tablet play plan with supervision from parents | • Children showed clear intentionality when planning their games, making choices without much assistance - children chose from the game apps available in their family's tablet |
| Hunter et al. 2014 | Cooperative virtual interaction impact on creative play and social engagement | Children played together open-endedly inside digital environments that they select, create, and arrange | • Children had strong preferences about what each scene should look like, and voiced numerous suggestions for content and activities |
| McKenzie et al., 2014 | Mobile-phone game impact on children's outdoor physical activity | Children played a treasure hunt game using their mobiles. Each player received a a map with a set of predefined clues and movement activities. Players interacted with gameplay by hunting for treasure locations (QR codes) with a set of predefined clues and movement activities challenges | • Most children found the level of difficulty to be between "just right" and "too hard"<br>• Children engaged in deciding about the most strategic ways of advancing to the next clue location within a treasure-hunt game |
| Saksono et al., 2015 | Collaborative exergame impact on children's and parent's physical activity | Children and caregivers engaged in physical activity to earn time and points of a videogame in order to complete space missions | • It increased awareness of opportunities for exercise and motivated those already aware of physical activity opportunities to decide to follow them up |

*Table 7 Summary of studies relevant to decision making*

### 5.1.3.1   *Offering choices of pathways towards set play goals (5 studies)*

The main feature promoting decision making was *offering choices of play-pathways towards set play goals*. The study by Saksono et al. [135] looked at child engagement in exergames. Within this game, 14 eight year-old children could choose their own physical activity (PA) to fuel games. Results indicate how the exergame made children make more PA decisions as demonstrated by children becoming more aware of opportunities for PA. In addition, the study by McKenzie et al. [116] made 14 five to twelve year-old children progress between the clues of an outdoor treasure-hunt game within a digitally mapped outdoor territory. The map showed different possible routes children could take. The results demonstrate that providing choices engaged children in following (and deciding about) different strategic shortcuts between clue locations. Also, in Hiniker et al.'s [133] study, 11 four to six year-old children were observed to use Plan&Play app to define the order and time of their play choices. This app engaged children in a step-by-step planning process based on the offer of choices, to choose games, game play order and length. Children engaged in making choices about their play time without much need of adult help.

Furthermore, the study by Bai et al. [113] compared 14 four to six year-old children's play behaviours when an augmented reality magic mirror device allowed them or not to choose the setting, characters, and emotions of their pretend play characters. Results show that children engage in making all these types of decisions in order to play the game (and that decisions tended to be more imaginative when confronted with open-ended or ambiguous prompts). Finally, Hunter et al. [120] allowed the customisation of virtual worlds by adding or modifying the background or objects available in the scenes where 12 six to twelve year-olds and their parents played. The results showed that child participants had strong preferences about what the scene should look like, and voice numerous suggestions for content and activities.

We also found that decision making made children think about a variety of developmentally relevant domains, such as physical activity or emotions [113] [135] and engage in important cognitive processes, such as strategic thinking and causal elaborations in the study by McKenzie et al. [116] and Bai et al. [113]. All these contents and processes were closely linked to the content of the choices children were invited to make. For example, in Bai et al.'s study [113], decisions about emotions of pretend play characters led to children practising their emotional understanding and causal thinking about emotions. In McKenzie et al.'s study [116] children practiced their strategic thinking when making decisions about the best treasure-hunt route. In Hiniker et al.'s study [133], children practiced their planning skills when deciding about games, order and length of their video-gaming. And in the study by Hunter et al. [120] children practiced their imagination when choices were less well defined (open-ended or ambiguous).

## 5.1.4   Forcing children's problem solving and physical activity (8 studies)

Problem solving can be considered as the move from an initial state to a goal state when the steps between are not apparent [163]. This domain of thought is considered a basic life function [164] and a core component of emotional intelligence [165, 166]. There is evidence that children can represent solutions to problems mentally from as young as 1-year-old [167]. Benefits of collaborative problem solving in particular, have been extensively documented (e.g.: [148, 168]).  On the other hand, physical activity refers to "any bodily movement produced by skeletal muscles that results in energy expenditure" (p.126) [169]. The benefits of physical activity are widespread and well documented in systematic reviews ranging from physical health [170] to cognition, emotion, and academic achievement [171, 172, 109, 108]. Children's levels of physical activity have been shown to decline with age [173]. Given its relevance for health, cognitive and emotional development, the study of why physical activity declines [174] and how to promote it among children has surged in the last decades. The latest evidence shows that current interventions are not proving to be as effective as desired for children [175, 176].

Within this review, forcing children's problem solving and physical activity refers to how play technologies were designed and programmed to require children to engage problem solving and physical activity during play. This was delivered by including such actions as requirements to achieve play goals or as requirements to engage in play activities. We found examples of how play technology did this for the cases of problem solving in the studies by Malinverni et al. [123], McKenzie et al. [116], and Shahid, Krahmer, and Swerts [117]; or physical activity in the studies by Cibrian et al. [136], Garde et al. [119], Martin-Niedecken [130], Saksono et al. [135], and Straker et al. [121]. Table 8 and Table 9 show a summary of these studies, including only results relevant to problem solving and physical activity, respectively.

| Author & year | Objective (To study) | Play Goals | Key findings (relevant to review results) |
|---|---|---|---|
| Malinverni et al., 2018 | Augmented & mixed reality impact on children's understanding, collaboration and use of physical space | Children played using either a shared tablet or portable projector. They found "magical portals" (markers) located on the walls of two different spaces within the school in order to solve a mystery | • Children engaged in problem-solving to solve a mystery when using the projector or the tablet |
| McKenzie et al., 2014 | Mobile-phone game impact on children's outdoor physical activity | Children played a treasure hunt game using their mobiles. Each player received a a map with a set of predefined clues and movement activities. Players interacted with gameplay by hunting for treasure locations (QR codes) with a set of predefined clues and movement activities challenges | • Most children found the level of difficulty to be between "just right" and "too hard"<br>• Children engaged in deciding about the most strategic ways of advancing to the next clue location within a treasure-hunt game |
| Shahid et al. 2014 | Social robot impact on children's enjoyment and expression | Children played a computer card game either alone, in collaboration with a friend or with a robot. When winning children collected coins as rewards | • Children engaged in problem solving to play the game* (children had to guess if an upcoming number was higher or lower than a previous number within a row of 6 cards) |

\* Information inferred by reviewers from the report of each study

*Table 8 Summary of studies relevant to problem solving*

| Author & year | Objective (To study) | Play Goals | Key findings (relevant to review results) |
|---|---|---|---|
| Cibrian et al. 2016 | Fabric-based interactive surfaces impact on children's development | Play with the interactive fabric (BendableSound) during free-play in their classrooms. Children use movement and touch to erease nebulas, reveal underlying space elements and play music | • Children improved in their motor skills |
| Garde et al., 2016 | Exergame impact on children's school-based physical activity | Children accumulated points for a mobile videogame by means of physical activity outside the videogame environment | • Physical activity increased in equal degrees among boys and girls<br>• After a week of washout period, physical activity returned to its baseline normality |
| Martin-Niedecken, 2018 | Motion controllers impact on children's social and bodily interplay | Children played a videogame using two types of body motion controllers: a full body motion controller (FBMC) or a Kinect. They operated in the world of a young pirate, searching for buried treasures with a flying ship and overcoming challenges using physical movements | • Participants engaged more in interdependent bodily interplay with the FBMC than the Kinect |
| McKenzie et al., 2014 | Mobile-phone game impact on children's outdoor physical activity | Children played a treasure hunt game using their mobiles. Each player received a a map with a set of predefined clues and movement activities. Players interacted with gameplay by hunting for treasure locations (QR codes) with a set of predefined clues and movement activities challenges | • In their self-reports almost all children (13) indicated they felt doing a lot of running and jumping during the game, and 5 of them considered themselves tired after it |
| Saksono et al., 2015 | Collaborative exergame impact on children's and parent's physical activity | Children and caregivers engaged in physical activity to earn time and points of a videogame in order to complete space missions | • Caregivers and children collectively assessed their physical activity reached during the game<br>• It increased awareness of opportunities for exercise and motivated those already aware of physical activity opportunities to follow them up |
| Straker et al., 2009 | Comparison between different types of screen interaction and impact on children's muscle activity | Children watched an animated film and played various video games using five different game devices | • Use of the wheel controller resulted in some increase in upper limb movement and muscle activity, but the other traditional input devices were usually as sedentary as watching a DVD<br>• Use of the active-input device based on body movements (EyeToy) resulted in considerably greater activity at all muscles |

*Table 9 Summary of studies relevant to physical activity*

## 5.1.4.1  Forcing problem solving (3 studies)

The study by Malinverni et al. [123] engaged 36 children aged nine to eleven in problem solving in a mixed-reality mystery solving game under two conditions: Tablet-based and a projector-based mixed reality. Here children were required to solve problems in order to make progress within the mystery solving game. Results from post-play interviews show that the children engaged in problem-solving in both conditions.

The study by McKenzie et al. [116], exploring 5-12 year-olds' play, also engaged children in problem solving. In particular, authors indicated that children engaged in deciding about the most strategic ways of advancing to the next clue location within a treasure-hunt game. Players themselves indicated that the level of challenge of the game tended to be between "just right" and "too hard", which also indicates the existence of certain levels of difficulty.

Finally, Shahid et al.'s [117] study looked at 8-12 year olds from two different cultures playing a computer card game across three conditions: alone, with a robot (the iCat), or with a friend. Children had to guess if an upcoming number was higher or lower than a previous number within a row of 6 cards. The focus of the research, however, was not problem solving (e.g., strategies children used to make decisions/guesses about the upcoming card). Instead, researchers studied children's emotional reactions when they failed during this problem solving play activity. We infer that children engaged in problem solving to play the game given the nature of the game.

### 5.1.4.2   Forcing physical activity (6 studies)

Play technologies also involved children in physical activity (PA) because this was central to engagement in play activities or necessary for action to achieve play goals. This was the case for exergames, studied by Garde et al. and Saksono et al., that programmed play in such a way that 28 nine to thirteen year-olds [119] and 14 eight year-olds [135] could not keep playing if they did not accumulate enough PA. Similarly, McKenzie et al.'s [116] outdoor treasure-hunt game required 14 five to twelve year-old children to move (walk or run) between clue locations marked with QR codes and overcome physical challenges to progress through the game. The same could be said to have been the case for the video-games controlled through EyeToy or FBMC/Kinect that detected whole-body movements by 20 nine to twelve and 32 ten to fourteen year-olds in the studies by Straker et al. [121] and Martin-Niedecken [130]. A similar case was found for the Bendable Sound screen studied by Cibrian et al. which was operated through gross body movement and hand touch studied among 22 two to three year-old children [136]. The games led to either an increase in PA [119, 121], engagement in vigorous PA [116], awareness of PA opportunities [135], or improvement in motor skills [136]. In all cases the interactive rules of the play technology made children move their whole bodies in order to fuel or control (i.e. play) the games. In some cases the specific type or amount of body movements encouraged depended not only on the specific play missions, but also the physicality of the material elements used to control or structure the games (FBMC, Bendable Sound, and treasure-hunt QR codes).

# 6   Principles underpinning the way developmentally relevant behaviours are promoted by play technologies

This systematic review aimed to gain a better understanding about the interactivity link that the literature has suggested drives the effects of digital play on child development. This review scrutinised how new phygital play technologies – technologies that engage children in analogue actions during digital play – promote developmentally relevant child behaviour. In relation to RQ1 (What type of social, emotional, cognitive or physical developmentally relevant actions do phygital play technologies afford for children?), we identify four types of play technology behavioural affordances: facilitating child self-monitoring; promoting collaboration; promoting children's decision making; and forcing problem solving and physical activity. With regards to RQ2 (How are these affordances delivered by phygital play technologies?), the review found that the different behavioural affordances driven by phygital play technologies were supported by an array of socio-technical partnerships between the technology and users. These partnerships emerged at the interplay between a variety of digital play mechanics, physical characteristics of technology and play goals pursued by children.

Following our *goal-oriented and tool-mediated action* theoretical framework, we further categorised the overarching principles that underpin the ways in which play technologies (tools) promoted developmentally relevant behaviours in children. We identified four affording principles: action regulation; social expectations;

technical features of phygital play technologies; and play goal-tool-action alignment. Table 10 presents a summary linking each of the afforded behaviours to affording forces related to these principles, such as play goals, action regulation, social expectations, play mechanics and physical characteristics of phygital play technologies.

| Developmentally relevant actions (DR actions) | Goals and Social expectations promoting DR actions | DR action regulation | Play mechanic affording DR actions | Physical features promoting DR actions |
|---|---|---|---|---|
| Self-monitoring | Children engaged self-monitoring to stick to play rules or advance through intermediate play goals | **Inviting action:** Technology providing information about own performance | **Most evidence** - 5 in 6 studies. Examples include: robot alerts to stop social conflicts; informing about collection of fuel for video-gaming; informing progress status of players to stick to children pace and strategise own treasure-hunt. | **Limited evidence** -1 in 6 studies. This could be seen in how the size and flexible material of a touch screen required children to pay attention to information about the appropriateness of own touch strength and balance in order to play. |
| Coordinated or collaborative joint play | Playing together is communicated as a desirable (expected) way to play | **Inviting actions:** Technology slowing down and pausing play pace | **All evidence** - 3 in 3 studies. Examples include: using slower pace or self-pace to make tablet games more socially inclusive. | **No evidence** |
| | | **Forcing action:** Technology making play interdependent | **All evidence** - 4 in 4 studies. Examples include: connecting the input of two controllers as if they were one; creating a platform for two people to interact physically mediated by camera feeds while in different places. | **No evidence** |
| | | **Inviting action:** Technology making objects of social play jointly accessible | **Some evidence** - 4 in 9 studies. Examples include: connecting two depth camera feeds into one projected image seen by both players; allowing for parallel multi-touch in tablet games. | **Most evidence** - 7 in 9 studies. Examples include: using projections rather than tablets to encourage collaboration in a group mystery game; making use of a large enough screen and physical materials to help children coordinate own pretend play through a magic mirror augmented reality device. |
| Decision making | Decision making as intermediate-goal to advance towards play goals | **Guiding action:** Technology offering choices of pathways towards set goals | **All evidence** - 5 in 5 studies. Examples include: providing a selection of pretend character emotions to enrich pretend stories; providing a series of possible outdoor paths in an outdoor treasure-hunt; giving freedom to select own physical actvity to fuel video-gaming time. | **No evidence** |
| Problem solving & physical activity | Engaging forced actions is required to achieve pre-set play goals or keep playing | **Forcing action:** Technology requiring developmentally relevant actions as means to play and reach game goals | **All evidence** - 8 in 8 studies. Examples include: making children engage in problem solving to advance in treasure-hunt games and a card guessing game; making children engage in physical activity by incorporating body movement and exercise as actions to fuel video-games, control video-games and move between clue points in an outdoor treasure-game. | **Some evidence** - 3 in 8 studies. Examples include: scattering clues of treasure-hunt or mystery games around physical spaces; spacing the control buttons of FBMC at different heights promoted jumping up and down; the flexibility, up-right position and large size of the Bendable Sound touch screen affording children's efforts to balance their bodies; larger size of shared screens/projectors facilitating joint problem solving. |

*Table 10 Socio-technical partnerships between phygital play technologies and child behaviour*

## 6.1 Action regulation

It is evident that the different types of affordances generated by play technologies could be said to vary in terms of their directiveness, that is, how much they invited, guided or forced developmentally relevant actions.

We suggest that the facilitation of self-monitoring could be seen as a case of *invited* action possibilities. The promotion of decision making could be considered to represent *guided* action possibilities. And the cases of problem solving and physical activity could be understood as examples of *forced* actions possibilities. *Invitations* were extended through delivering mechanisms such as providing information about own performance for self-monitoring; slowing or pausing play pace; ensuring the accessibility to joint attention for collaboration; and freeing the array of possible decisions (strategic choices) one could make to advance within games. Invitations commonly made developmentally relevant actions more attractive (easily achievable and convenient) for children's achievement of play goals. *Guiding* included the mechanism of offering players specific choices of pathways towards set goals to promote decision making. Guiding mechanisms reduced the array of children's possible actions and directed attention towards such choices and choice-bound actions. *Forcing* included mechanisms such as making play interdependent to promote collaboration, and using problem solving and physical activity as required means to play. Forcing mechanisms commonly limited the type of actions children could engage to play.

Our review findings are consistent with play literature suggesting that play can generate a zone of proximal development for children [39]. In particular, it suggests that play technology can regulate children to engage in activity, thinking, and understandings that they might not engage otherwise [48]. Furthermore, our findings are also consistent with previous theories indicating that people's behaviour can be both constrained and forced by features of technology [87, 93]. This review makes a contribution, however, in helping to further specify levels of directiveness of such constraints as either inviting, guiding, or forcing.

## 6.2   Social expectations guiding behaviours

Play rules also guided behaviour within play. This was the case of collaboration, when researchers introduced the expectation that children would play together. Social expectations become a sort of procedural goal for children to bear in mind when playing. We observed this when children were explicitly asked to play together, for example in the studies by Bai et al. [113] or Cohen et al. [126]. We also observed this when children were left or asked to share one play device in the studied by Lawrence [134], Hiniker et al. [132], Malinverni et al. [123] and Cibrian et al. [136]. The relevance of social rules or expectations within play technology expands our original theoretical framework. Digital play theory tends to consider rules embedded as part of play mechanics to be the main type of rule affording player's behaviour [87, 88]. This review invites to include social rules or social expectations also to promote developmentally relevant actions within play technologies. This is in line with principles from activity theory that indicate how activity settings can guide actions [86].

## 6.3   Technical features of technology

While social expectations invited children's engagement in particular developmentally relevant actions, what actually made it more likely for children to engage such actions were the technical features of play technologies. In this section, we separate the features between digital play mechanics and physical characteristics of play technologies. As can be seen in Table 11, digital play mechanics were found to be more important for the case of self-monitoring, decision making, and problem solving. And both digital play mechanics and physical characteristics of technologies were found to be as relevant for collaboration and physical activity. In the following sub-sections we indicate how these two features of phygital play technologies related to the developmental behaviours afforded in the studies reviewed.

| Author & year | Self-monitoring | Collaboration - pauses/speed | Collaboration - Interdependence | Collaboration - joint attention | Decision making | Problem solving* & Physical activity✛ |
|---|---|---|---|---|---|---|
| Andrist et al. 2013 | | | PM | PhF | | |
| Bai et al. 2015 | | | | PM & PhF | PM | |
| Cibrian et al. 2016 | PhF | | | PhF | | PM ✛ |
| Cohen et al. 2014 | | | PM | | | |
| Garde et al. 2016 | PM | | | | | PM ✛ |
| Hiniker et al. 2017 | PM | | | | PM | |
| Hiniker et al. 2018 | | PM | | PM & PhF | | |
| Hunter et al. 2014 | | | PM | PM & PhF | PM | |
| Lawrence 2018 | | PM | | | | |
| Malinverni et al. 2018 | | | | PhF | | PM * |
| Martin-Niedecken 2018 | | | PM | PhF | | PM & PhF ✛ |
| McKenzie et al. 2014 | PM | | | | PM | PM & PhF * ✛ |
| Saksono et al. 2015 | PM | | | PM | PM | PM ✛ |
| Shahid et al. 2014 | | | | | | PM* |
| Shahid 2018 | | | | PhF | | |
| Shen et al. 2018 | PM | PM | | | | |
| Straker et al. 2009 | | | | | | PM & PhF ✛ |

Legend: PM = Play Mechanic ; PhF = Physical Feature

*Table 11 Technical features of technology affording developmentally relevant actions*

## 6.3.1   Technical features facilitating child self-monitoring

Digital play mechanics providing performance information for self-monitoring included: robot alerts to stop social conflicts in Shen et al.'s study [118]; providing information about own collection of PA to fuel a video-game in Saksono et al.'s [135] and Garde et al.'s [119] studies; displaying information about own progress through own self-set play plan in Hiniker et al.'s study [133]; and giving information of progress status of all players to enhance awareness of own pace and route choices in a treasure-hunt game in the study by McKenzie et al. [116]. Physical characteristics of technology generating information about own performance were much less frequent. Within this review the only example found was that of Cibrian et al.'s Bendable Sound platform and its physical characteristics [136]. In particular, this device promoted self-monitoring of own balance and strength. It made children interact with a flexible fabric touch screen which provided immediate information about the appropriateness of players' strength and balance to control the video-game through touch. It is important to note, however, that these physical characteristics of the Bendable Sound would not have been relevant for the promotion of self-monitoring without the interactive features (digital play mechanics) of the video-game played.

### 6.3.2   Technical features promoting collaboration

Collaboration was more likely to occur when digital play mechanics allowed for more self-paced, slower play, or even paused play in tablet games, according to the findings reported by Lawrence [134], Hiniker et al. [132] or Shen et al. [118]. Digital play mechanics of interdependency between players also helped collaboration, such as in the case of using interconnected full body motion controllers in Martin-Niedecken's study [130], or depth camera feeds to play as designed by Cohen et al. [126] and Hunter et al. [120]. To promote collaboration, however, interdependency needs to be complemented with assurance to joint accessibility of play objects. This indicates the importance of physical characteristics in phygital play technologies to promote collaboration. This was evidenced by the more collaborative nature of Full Body Motion Controller (with physical buttons) versus the smart eye Kinect [130]. The importance of the physical characteristics of phygital play technologies was also shown by the series of studies finding that projection of images on walls or large screens aided negotiation in play [115, 113]. Therefore, we can see that the features that afforded collaboration depended on both the digital play mechanics as well as physical characteristics of play technologies.

It is important to note, however, that, interdependency aside, the effect of these digital play mechanics and physical characteristics would not have instigated collaboration if the games they supported were not framed as social games. Playing together became a social expectation, hence a procedural goal that children tried to achieve/comply with. Evidently, a different behaviour would have been practiced by children if the rules of the game were to play in parallel or play against each other rather than working together when using the technology.

### 6.3.3   Technical features inviting children's decision making

The results indicate the importance of digital play mechanics rather than physical characteristics of play technologies in promoting children's decision making. Each play environment offered choices of pathways towards set goals. Sometimes choice making was explicitly prompted by interactive features, such as choosing between different emotional expressions for pretend play characters in the study by Bai et al. [113] or between different games, order and length of game play in that by Hiniker et al. [133]. Other times, choice making was prompted implicitly by the demarcation of play limits. Within such demarked limits, children could take decisions about strategies to make beneficial gains within games. This was the case when children were left to decide the type of physical activity (PA) they would engage to fuel a video-game in Saksono et al.'s study [135], or when left to decide the paths they could take in an outdoor treasure-hunt game in McKenzie et al.'s study [116]. In the case of Bai et al.'s [113] and Hunter et al.'s [120] studies, the children would decide on the play environment itself, what scenes to play in and share with their parents or friends, or what objects to interact with.  In either case, what invited (and allowed) children to make decisions were the causal (if-then) interactive rules of games (e.g., if children decided about a good PA, then they could get a good amount of video-gaming time).

### 6.3.4   Technical features forcing problem solving and physical activity

Features promoting problem solving and physical activity (PA) through play technologies are based on digital play mechanics and physical characteristics of play technologies. Makers programmed the need for children to engage in these developmentally relevant play actions as an integral part of game experiences, to play or advance within the games. Shahid et al. [117], Malinverni et al. [123] and McKenzie et al. [116] offer examples of digital play mechanics that enforce children to engage in problem solving by requiring them to solve problems to advance in treasure-hunt games and a card guessing game. Greater physical accessibility of digital content (visibility of projections rather than tablet screens) also made problem solving more likely to happen within group contexts in Malinverni et al. [123].

Digital play mechanics made children engage in PA by incorporating body movement and exercise as a key action to fuel video-games in the studies by Garde et al. and Saksono et al. [119, 135], control video-games in the studies by Martin-Niedecken and Straker et al. [130, 121], and move between clue points in treasure-hunt games in Malinverni et al.'s [123] and McKenzie et al.'s [116] studies. The physical characteristics of play technologies also had a role in the facilitation of these developmentally relevant actions in some play technologies. For example, in Malinverni et al. [123] and McKenzie et al. [116], treasure-hunts games required clue spots to be scattered along large outdoors or indoors areas.

Similarly, the physical characteristics of Full Body Motion Controller (FBMC) or the Bendable screen made children engage in particular types of physical activity, such as jumping or balancing. Jumping up and down was afforded by spacing the control buttons of FBMC at different heights of children in Martin-Niedecken's study [130]. And whole-body balance practice was afforded by the high flexibility and large size of the Bendable Sound touch screen in Cibrian et al.'s study [136].

Consistently with Norman [93] and Salen and Zimmerman [87], our analysis demonstrates that both digital play mechanics and physical features of phygital technologies can promote child behaviour. This review adds to the literature, however, in indicating that, in general, new phygital play technologies tend to promote child behaviour more through digital play mechanics than physical features of technology.

## 6.4   Play goal-tool-action alignment

From our analysis we also hypothesise that play goal-tool alignment could well be thought to be the *structure of action possibilities* affording specific developmentally relevant actions. Actions complete a sort of in-waiting harness generated together between goals and tools. Children's play actions then transform this harness into a whole goal-tool-action alignment in practice.

The importance of *goal-tool-action alignment* within play could be seen more clearly in the experimental studies reviewed. These studies compared the effect of changes in play technology over child behaviour while keeping play goals fixed across conditions. The experiment by Bai et al. [113] on the magic mirror used for social pretend play is a good example. In one experimental condition the technology offered children choices of emotions for their pretend play characters. In the other condition the technology did not offer such a choice. All other features of play were equal across conditions. As previously indicated, the results showed that children referred much more to emotional states, played more pretend acts involving emotional states, and offered more causal elaborations of characters' emotions when character emotions could be chosen. Therefore, emotions became another feature in relation to which children could pretend, an extra vehicle to achieve richer pretend play stories. Consequently, the addition of the tool of emotion selection in a situation where children were engaged in pretend play as a goal, led children to engage in all these emotions related elaborative behaviours (actions). Moreover, if the goal of the play situation had been different, such as, for example, to choose an emotion that represented better the player's emotional state, none of these emotional elaborations would have taken place.

The study by Shen et al. [118] comparing conflict resolution between children moderated by Keepon robot also illustrates well the importance of *goal-tool-action alignment* for the developmental affordances generated by play technology. The authors compared two conditions. In the first condition Keepon directed different aspects of play but did not do anything when children entered into social conflicts. In the second condition Keepon directed different aspects of play and also flagged when it detected a social conflict and suggested ways in which children might be able to resolve the problem. Children were four times more likely to resolve conflicts positively under the latter condition. Therefore, the technology changed children's behaviour by introducing a new tool (identification of and guidance for social conflicts) and goal (to play without conflicts), and making sure that these were aligned to harness behaviour.

Our results are consistent with previous theory relating actions to goal-oriented tool mediation [85, 86]. Previous analytical works have recently indicated the relevance that tools and goals can have for actions engaged for learning in serious games [96]. This review shows that tools and goals can also be relevant to understand links between play technologies and child behaviour when engagement is motivated by enjoyment rather than learning.

# 7  Conclusion

We embarked on this review with two key research questions, which we now turn back to address.

***RQ.1 What type of social, emotional, cognitive or physical developmentally relevant actions do phygital play technologies afford for children?***

From our analysis of the higher quality papers we can conclude that, overall, different types of developmentally relevant behaviours were afforded by play technologies. Play technologies afforded actions that were of cognitive, social, and physical nature. In particular, these actions included: 1) facilitating child self-monitoring; 2) promoting collaboration; 3) inviting children's decision making; and 4) forcing problem solving and physical activity. There were also some emotional aspects afforded, but these did not emerge as themes (e.g., casual elaborations of emotions).

***RQ.2 How are these affordances delivered by phygital play technologies?***

We identified a series of specific characteristics of the interactivity of play technologies that promoted developmentally relevant child behaviour. Self-monitoring was promoted by the provision to players of information about their own performance. Collaboration was promoted by slowing down and pausing play interactivity, making play interdependent, or making objects of social play jointly accessible. Decision making was invited by offering choices of pathways towards already set play goals. Problem solving and physical activity were forced by using these types of actions as necessary means to keep playing.

Based on these specific characteristics of phygital play technologies promoting developmentally relevant child behaviour, we further identified four overarching principles underpinning the ways in which phygital play technologies and play situations afforded child behaviour: i) action regulation; ii) social expectations; iii) technical features such as digital play mechanics and physical characteristics of technologies, and; iv) goal-tool-action alignment. These principles should be considered to reflect how the *goal-oriented and tool-mediated action* framework is put to work to explain how the play interactivity of new phygital play technologies promotes children's developmentally relevant behaviour.

First, action regulation indicated that the child behaviour afforded could be increased differently through different degrees of directiveness. That is, technologies could invite, guide or even force child behaviour. Second, social expectations played an important role in affording child behaviour. This was particularly evident for the case of collaboration. When researchers framed the play situation as social play children were likely to engage with others (if the technology permitted). Third, digital play mechanics, or the action rules programmed into phygital play technologies, were very important to afford children's play behaviour. We found strong evidence of their effect on all developmentally relevant behaviours studied across the reviewed studies. Physical characteristics of technologies were also found to afford children's play behaviours but only for some developmentally relevant behaviours. They were particularly relevant when promoting collaboration (e.g., sizes of screens) and physical activity (e.g., scattering clues in the outdoors to encourage walking or running, placing push buttons at different heights to encourage jumping and squatting). And fourth, goal-tool-action alignment indicated that child behaviour was generally afforded by making specific types of actions the most convenient or intuitive way of fulfilling goal-tool structures of action possibilities. When children used different tools to achieve the same goals their actions changed.

# 8  Summary, implications and future work

Overall, the review leads us to conclude that new play technologies engaging children in digitally enhanced analogue behaviours can be beneficial for development among typically developing children. We found that the interactivity of these new technologies has the potential to invite, guide, or force developmentally appropriate child behaviour. They do so by framing desirable behaviours as missing pieces that children need to activate to complete the whole formed between digital or physical features of technology and play goals/expectations. In practice, the physical or digital characteristics of new play technologies can be designed and programmed to function as aids that help children achieve their play goals through targeted (expected) developmentally relevant behaviour.

In the future, technology designers and programmers might benefit from using the insights from this review to enhance the effect that their new and improved play technologies can have on child development. At the more empirical and practical level, the developmentally relevant actions identified (self-monitoring, collaboration, decision making, problem solving and physical activity) and the different ways these behaviours were found to be delivered by play technology affordances (e.g., contingent feedback about own performance, slowing interactivity) can provide achievable targets and concrete guidelines for design of interactive phygital play technologies. These elements are not prescriptive but do provide a theoretically and empirically grounded basis on which to build. They introduce a way for child-computer interaction researchers to think about how they frame the type of developmental affordances they are finding in their own studies. For example, Kirginas et al. [177] clearly showed the benefits of free-form play, but these could have been framed and discussed in terms of inviting children's decision making (as opposed to guiding or forcing them). Parsons et al. [178] explored the area of collaboration in great depth, but additional benefits identified in this review, such as decision making and problem solving, were potentially present yet not explicitly considered.

Future research could expand the exploration of the effects of phygital play to other transferable skills too. In particular, the present review evidenced a clear need to carry out more studies of phygital play technologies in relation to social and emotional skills, such as emotion regulation, emotion understanding or awareness, theory of mind, negotiation skills, conflict management skills and general social competence. More cognitive skills could also be studied, such as executive functions, reasoning, abstract thinking and creativity. All of these have been studied to some extent in relation to different types of analogue play [179, 180], with some of them (social competence, emotional awareness, theory of mind, and executive functions) showing stronger evidence of developing from play [180].

Similarly, in the more pedagogically focused work carried out within the child-computer interaction community, parallels could be extended in future research between phygital play technologies and other playful learning spaces such as educational makerspaces. The makerspace movement is rooted in constructivist, student-centred, interest-driven educational theories, using technological tools for 'making things', such as physical computing kits powered by virtual programming [181]. Studies suggest that early year engagement in makerspaces enables the development of individual agency, fosters social interaction and allows children to transition seamlessly across digital and non-digital domains in their maker play [182]. This final aspect has the potential to shape institutional pedagogical practices, coined in the literature as 'postdigital play' [183]. Whilst the focus of this review was restricted on play purely for enjoyment purposes, future research would benefit from trying to apply the concepts developed in it to research on makerspaces and postdigital play.

Additionally, in terms of theory, the review responds to recent calls within the child-computer interaction community to develop more intermediate-level knowledge such as 'strong concepts' within the field. That is, concepts carrying core design ideas, generally derived from studies on specific designs, that reside in the interface between technology and people [184]. The strong concepts we identified refer to the principles underpinning the way in which phygital play technologies afforded child behaviour: action regulation, social expectations, technical features of technologies, and goal-tool-action alignment. These concepts can be used as analytical tools to design or evaluate phygital play technologies by anticipating or making sense of their 'behavioural success'. For example, designers might want to determine the level of action regulation for which they desire to design (inviting, guiding, or forcing behaviour). These levels are likely to be consistent with their own perspective on the role of children in technology design [185, 51, 186] or the value they give to open-ended, guided, joint or even adult-directed play for child development [54, 187, 188, 189]. Designers can also reflect on the extent to which their device communicates social expectations that promote children's up-take of target behaviours, or think about the physical and mechanical features of the technological solutions that afford such behaviours. This could be done either before or after technology trials, depending on the interest that designers have in carrying out designer-led, child-led, participatory or cooperative technology design [51, 190].

Similarly, the concept of goal-tool-action alignment can be used to reflect on the level of anticipated or observed alignment between play artefacts, play goals and target behaviours. These concepts might be of particular interest to practitioners and scholars working in technologies bridging the physical and the digital, such as those promoting full-body interactions [191]embodied child-computer interactions [192] or the use of tangibles [193] within play. For instance, the goal-tool-action alignment analytic could have been used to further strengthen the work by Superti Pantoja et al. [194]. The authors developed and evaluated tangible voice agents (Wizard-of-Oz operated) to promote social pretend play and socio-dramatic play. They used a more open approach to develop the technology, guided by their 3-4 year-old participants' engagement with the tangibles. The technology worked well when children were not distracted by using tablets themselves to control the voice agents. Using the goal-tool-action analytic to interpret the outcomes and affordances of the technology at each iteration of the design journey may have simplified the process (e.g., by helping them anticipate - based on previous research - that the inclusion of the tablet would have afforded behaviours unrelated to pretend or socio-dramatic play). Also, analysing and reporting the results using this analytic would have provided general insights more easily transferable to other settings or types of social play, a limitation identified by the authors themselves.

Although we do recognise many complexities, challenges, and value considerations inherently involved in the design decisions of play technologies, we argue that more intentional considerations of the potential developmental benefits highlighted in this study can lead to more developmentally supportive play technologies. Designing with more intentionality in mind calls to embrace, to some degree, the structuring effects of artefacts in human behaviour [77]. That does not mean to fall into deterministic visions of technology; children will continue giving their own meaning and use to digital play artefacts [195], and artefacts can and will continue to be designed to help children play openly [54]. What we mean is that, starting to think about play technology more as a tool to deliver 'vicarious and unnoticeable adult guidance' [196] of play behaviours could help children develop towards skills valued by society whilst playing. This perspective is in line with decades of research about the importance of scaffolding or guidance for development [40, 41, 197, 198, 199, 200] as well as increasing recent evidence about the benefits of guided (child-led but adult-assisted) play for development [189, 35] or free play facilitated by unintrusive adult structuring [201].

Moreover, future research on play technologies may benefit from using the theoretical framework specified here to evaluate the effects of new phygital play technologies as they evolve and spread among the general and typically developing child population. For example, new studies would benefit from comparing the benefits of interactivity inviting, guiding or forcing developmentally relevant actions through phygital play technologies on child behaviour and subsequent development. Researchers might also want to compare the behavioural effects of framing the goals (purpose) of the use of phygital play technology as openly educational versus as tools for play among children. It would also be beneficial to keep exploring the importance of physical features of digital technology for developmentally relevant child behaviour (e.g., in relation to self-monitoring, decision making, and problem solving). It is important to note that, due to the relatively small number of studies involved in this review and the fact that many of these studies were not deliberately carried out to study child development, future research would also benefit from more empirical work to further test and verify the theoretical framework developed and used in this study.

Furthermore, although this review found that digitally enhanced play behaviours can have a positive effect on child behaviour, questions about the magnitude of effect of phygital play technologies in comparison to traditional analogue play, remain to be answered. The scarcity and array of outcomes studied by the quantitative studies reviewed made it impossible to address such a query. Additionally, we only focused on studies carried out with typically developing children and, with one exception, the studies selected for the review were all carried out with children living in Western contexts. Therefore, questions about the way in which phygital play may or may not benefit more diverse children, such as children from the Global South, neurodiverse children and children with physical disabilities are still to be addressed. The same could be said in relation to gender, ethnicity or culture, and other socio-cultural characteristics of children, which we found are largely ignored in the extant evidence about phygital play and developmentally relevant behaviours. Therefore, we would like to encourage researchers to undertake more carefully designed and intentional studies to address some of these questions. This would enable future specialist review synthesis to reveal the type of transferable skills and physical activity for which phygital (as opposed to fully digital and fully analogue) play may be better suited to promote through childhood in general and in more diverse groups of children in particular.

## Acknowledgements

In memory of David Whitebread (founder of PEDAL Centre, University of Cambridge) for all his contributions to research on play and child development; an inspirational academic guide and friend of the main author of this review. This study was funded by EPSRC (EP/P025544/2).

# Appendices

Appendix A

**Key information extracted from papers**

All papers

1. Research objectives, questions, and/or hypothesis
2. Type of technology studied
3. Key functions of the technology that were available to users (i.e. what does the technology do for the users?)
4. Developmental outcomes or developmentally relevant behaviours promoted by the technology (as concluded by authors)
5. Characteristics of the technology that explained developmentally relevant outcomes or behaviours (as suggested by authors)
6. Information about adult offers of guidance or directions to children during play sessions (as delivered rather than planned by authors)

Quantitative studies

1. Results from statistical analyses
2. Characteristics of the group conditions (those to which participants were exposed), if an experiment
3. Integrity of the intervention implementation (was it implemented as intended across experimental groups?), if an experiment

Qualitative studies

1. Results (e.g. Themes) related to cognitive, social, emotional or physical development /behaviours (as indicated by authors)
2. Quotes/discourse/descriptions from or about participants used to illustrate results
3. Characteristics of the study context considered to be relevant for outcomes according to authors (e.g., patterns within the setting of the study)

Appendix B

**Other descriptive information extracted from papers**

All papers
  1. Sample and its characteristics
  2. Setting of the study
  3. Methods applied

Quantitative papers
  1. Measures (variables, associated reliability, measurement times)
  2. Length of intervention (if applicable)

Qualitative studies
  • No extra fields extracted

Appendix C
**Checklists used for assessment of quantitative and qualitative studies**

CASP Checklist for assessment of qualitative evidence (adaptations in italics)

1. Was there a clear statement of the aims of the research? (before reporting results) – (*Can reviewers articulate the aims and the relevance of the study?*)

2. Is a qualitative methodology appropriate?

3. Was the research design appropriate to address the aims of the research?

4. Was the recruitment strategy appropriate to the aims of the research?

5. Was the data collected in a way that addressed the research issue? (*rationale for methods*)

6. Has the relationship between researcher and participants been adequately considered? (*either through study planning or reflecting back on the study*)

7. Have *ANY* ethical issues been taken into consideration?

8. Was the data analysis sufficiently rigorous?

9. Is there a clear statement of findings? *(throughout findings)*

10. How valuable is the research? *(have the authors discussed the study's value)*

Checklist includes guidance for reviewers. Full checklist is available online: https://casp-uk.net/wp-content/uploads/2018/03/CASP-Qualitative-Checklist-2018_fillable_form.pdf (Free open access).

Downs and Black [111] Checklist for the assessment of the methodological quality both of randomised and non-randomised quantitative studies (adaptations in italics; original criterion numbering shown as "D&B#")

1. Aim: Is the hypothesis/aim/objective of the study clearly described? (*regardless of consistency between aim and reported study*) (D&B 1)

2. Clear dependent variable: Are the main outcomes to be measured clearly described in the Introduction or Methods section? (*only 100% of Main DVs is assessed as yes*) (D&B 2)

3. Findings: Are the main findings of the study (*those discussed*) clearly described? (D&B 6)

4. Blinding: Was an attempt made to blind those people measuring the main outcomes *to either the hypothesis or the group allocation*? (D&B 15)

5. Measures: Were the main outcome measures used accurate (valid and reliable)? (D&B 20)

6. Comparability: Is there a risk that the contexts of the compared conditions might differ in any systematic way beyond the tested/manipulated play variable?  (only for experiments) (*added criterion*)

7. Randomisation: Were study subjects randomised to intervention groups? (*only applied for experiments*) (D&B 23)

8. Statistics: Were the statistical tests used to assess the main outcomes appropriate? (D&B 18)

9. Confound control: Was there adequate adjustment for confounding in the analyses from which the main findings were drawn? (*Unable to determined is granted if no confound is identified*). (D&B 25)

10. *Control of adult effect*: Is there a risk of adult-child interaction effects being confounded with independent variable (intervention) effects? (both role of adults and assignment of particular adult(s) to run different conditions are considered). (*added criterion*)

11. Representativity: Were the subjects asked to participate in the study representative of the entire population from which they were recruited? *Participants would be representative if they comprised the entire source population, used systematic sampling, stratification technique, or a random sample* (D&B 11)

Downs and Black developed their scale to assess interventions studies. To extend their tool to other types of quantitative studies, the criteria of Comparability and Randomisation were considered to be non-applicable (indicated as "NA") for some studies. Full checklist can be found here: https://jech.bmj.com/content/52/6/377 (Free open access).

# Appendix D

## Summary of higher quality studies reviewed

| Author & year | Objective (To study:) | #Children | Age | #Adult | Methodology | Play Time | Play Tool | Study Play Setup | Play Goals | Key Play Tech Features | Variables studied | Main findings (relevant to review RQs) |
|---|---|---|---|---|---|---|---|---|---|---|---|---|
| Andrist et al. (2013) | Virtual agent impact on children's turn taking, overlapping speech and fun | 33 | 4 -> 10 | 0 | Quantitative • Within subject experiment | Not stated | "Robo Fashion World" virtual agent computer app | Children played in groups of two to four. Game was presented on an LCD TV screen. Virtual agent presented the game and facilitated children's turn taking. | Players used an interactive language-based game to change the appearance of a model by calling out names of items on a game board, mediated by virtual character. | • use of gaze, proxemics and verbal interrogation to give cues to children • agent autonomously selects a sequence of actions to express to children • wizard as speech recogniser indicating to the system what children say | • Number of turns taken by each participant • Amount of overlapping speech | • The more flexible virtual agent able to apply actively all four types of turn-passing moderation strategies (gesture, gaze, proxemics and speech) was found to lead to more equal participation during play • Children's talk tended to overlap regardless of the type(s) of the virtual agent's moderation strategy |
| Bai et al. (2015) | Augmented reality system impact on children's joint pretence, emotion expression, divergent thinking | 14 | 4 -> 6 | 0 | Quantitative • Between subjects experiment • Observations | 15 min sessions | "FingAR Puppet" magic mirror augmented reality tabletop app | Children played in pairs with the magic mirror AR, situated in an in-between classrooms space. During the session, the experimenter and the teacher provided minimal prompts. In one condition children could choose the emotion of their play characters and in the other they could not. | Players created stories manipulating physical referents or shapes and assign functions/meanings to them (e.g., pirate), props (e.g., bicycle) and scenary textures/colours (e.g., grass). | • enables children's interaction with physical objects: puppets, blocks, shapes • enables children's selection of AR elements to create stories • enables children to change puppet's facial emotional expression • children can change role, prop or scenery | • Children's frequency of emotional state expression • causal elaboration of emotional state • explicit verbal communications of object transformations | • The AR system elicited large quantities of pretence play across all children • Children engage in a variety of types of communication to coordinate social sympretend play • Children were found to make more verbal communications of their transformations to playmates when choosing from open-ended representations (e.g., black) rather than from more definite-meaning representations (e.g., pirate) • Children generated more imaginary representations with familiar than with less familiar scenary textures, but when working with less familiar materials they tended to be more novel in their imagination • Children made explicit and deliverate decisions about their play when presented with options about characters and play backgrounds - they would also change the emotion of their characters if given the choice |
| Cibrian et al. (2016) | Fabric-based interactive surfaces impact on children's development | 22 | 2 -> 3 | 5 | Qualitative • Observations • Interviews | Sessions over 16 hours | "BendableSound" interactive projector Kinect app | Free-play in the classrooms. Children could play alone or with other children. | Players engaged in using their touch to erease nebulas to reveal underlying space elements and also to play music through their movement. | • tap, touch or grasp fabric canvas to remove obstructions • touch / move randomly appearing objects • tech responds to touch through sound | • Children's play behaviours • Skills related to age-relevant motor, cognitive, social and emotional development | • Children preferred to play together rather than alone (especially when creating sounds together) or when imitating each other / taking turns • Children improved in their motor skills • Children improved their sustained attention • Children improved communication skills |
| Cohen et al. (2014) | Video conference system impact on children's nature of distance play interactions | 9 | 6 -> 10 | 5 | Mixed-methods: • Within subject experiment • Observations | 10+ min sessions | "OneSpace" video-conference projector app Skype-like video-conference computer app | Participants played in pairs with props and toys. Most pairs (5) were parents with their children, and a few (2) were child-child pairs. Playmates were separated between two different room spaces. Participants played together using Skype and then using the OneSpace. OneSpace merged their full body personal images into one projected image. | Players played open-endedly together. | • merge video feeds of two remote sites into a shared visual scene • detection and display of person closest to respective camera | • Types of play • Nature of engagement between participants • Levels of physical activity | • The video conference system promoted more engagement in organised and cooperative play, whereas the conventional configuration promoted more parallel and associative play • The video conference system promoted more active/physical play, discussion and action between players, whereas the conventional configuration promoted more show-and-tell and make-believe play • In the conventional setting parents tended to coach children more and they would usually disengage with one another devolving into parallel or associative play or stopped playing |

| Author & year | Objective (To study:) | #Children | Age | #Adult | Methodology | Play Time | Play Tool | Study Play Setup | Play Goals | Key Play Tech Features | Variables studied | Main findings (relevant to review RQs) |
|---|---|---|---|---|---|---|---|---|---|---|---|---|
| Garde et al. (2016) | Exergame impact on school-based children's physical activity | 28 | 9 -> 13 | 0 | Qualitative • Within subject crossover experiment • Analytics | 15min sessions over 4 weeks | Exer-videogame smartphone app | Children played the exergame in teams (remotely). The study had two conditions: when collecting and when not collecting video-game time through accelerometers. Children could play whatever they wanted during the weeks of the study. | Players accumulate points through physical activity to use within a videogame. | • gold rewards for progression • collaborative points accumulation between teams • social messaging • rewards for physical activity | • Steps and length of active physical activity | • Physical activity increased in equal degrees among boys and girls • After a week of washout period, physical activity returned to its baseline normality • The exergame helped children monitor the amount of physical activity collected to fuel their videogame, hence incetivicing more physical activity to fuel play * |
| Hiniker et al. (2017) | Technology impact on children's self-regulation and parents' support | 11 | 4 -> 6 | 11 | Qualitative • Observations • Interviews | 3 min sessions | "Plan&Play" tablet app | Children played individually in their tablets. Parents were standing by to guide them in the planning and the use of the app. | Players aimed to plan and follow their own tablet play plan with supervision from parents. | • self-select games, order of play and time of play • parent-approved game plan • tracking and display of game plan status | • Children's intentionality in planning • Parents' attempts to scaffold children's interactions • Children's self-regulation | • Children showed clear intentionality when planning their games, making choices without much assistance - children chose from the game apps available in their family's tablet • Children self-regulated successfully, keeping to their plans in 93% of planned transitions • Most children required the support of their parents to understand how to use the app, but 88% could plan a second session without adult help • Both parents and children treated the app as a third party authority |
| Hiniker et al. (2018) | Traditional toys and digital apps impact on child-parent play | 15 | 4 -> 6 | 15 | Qualitative • Within subject experiment | 15 min sessions | Different tablet apps Traditional toys | Families brought a tablet with the child's favourite games & toys in the lab. The same children were observed playing with apps v. toys. Parent-child dyads chose whether to play together or not. | Players played with their favourite app games as well as played open-endedly with their favourite non-digital toys. | • various unspecified individual app games • assortment of unspecified non-digital toys | • Parent-child play behaviours, mainly in terms of social engagement and attention management | • Dyads were more likely to engage in joint activity when playing with toys than with tablets • Children tended to put toys, but not tablets, in the attentional spaces shared between them and their parents • Parents were more likely to engage children in turn taking and conversation when playing with toys • Tablets made difficult for parents to see what their children were doing, except when games allowed for multi-touch, which made children share the screen with parents to play together • Children tended to ignore parents' questions when playing tablet games • Children were more likely to trail off from conversations when app games showed visual effects or prompted them • When apps allowed for self-paced activity children were more able to manage their own attention • Traditional toys made children focus more on the play experience while also sustaining conversation with parents |
| Hunter et al. (2014) | Cooperative virtual interaction to support creative play and increase social engagement of geographically separated families | 12 | 6 -> 12 | 12 | Qualitative: • Interviews Quantitative: • Within subjects experiment • Observations | 2 X 90 min sessions | "WaaZam" video mediated communication system | 12 adult/child pairs in separate spaces with physical props such as puppets and toys. All went through four conditions: skype; merged personal windows under one player's ordinary background; merged personal images within digital play backgrounds created by artists; and merged personal images within digital environments customised by players. | To play together open-endedly as players are projected inside digital environments that they select, create, and arrange. | • conventional videoconferencing • a merged "magic mirror" mode • constructed fictional environments • merge video feeds of two remote sites into a shared visual scene • players can use existing scenes, customise scenes with additional objects, or create entirely new scenes | • Type of play • Adult/child play engagement • Adult/child mutuality • Behavioural characteristics | • Being together in the same virtual space was found to be the most effective way of making players engaged in shared activities and being creative. It also tended to increase play engagement and the diversity of play types • Participants focused more on each other when playing under a "mirror" mode • Personalisation appears to foster feelings of ownership, and can increase the richness and depth of play activities • Children had strong preferences about what each scene should look like, and voiced numerous suggestions for content and activities |

| Author & year | Objective (To study:) | #Children | Age | #Adult | Methodology | Play Time | Play Tool | Study Play Setup | Play Goals | Key Play Tech Features | Variables studied | Main findings (relevant to review RQs) |
|---|---|---|---|---|---|---|---|---|---|---|---|---|
| Lawrence (2018) | Collaborative play using tablets impact on children's interactions | 20 | 5 | 0 | Qualitative • Observations • Interviews | 15 min sessions | Different tablet apps | Children played a selection of 5 game apps with designated peers on a shared iPad in their classrooms. | Players play the games of choice. | • five literacy, numeracy and shape/size recognition apps • one open-design colouring app | • Types of play engaged by children • Children's collaborative play behaviours | • Children struggled for tablet control, especially in closed-ended reward based games and towards the beginning of play • Children cooperated, collaborated and enjoyed playing in the open-ended play app • Fast paced games turned players competitive, but slower paced games and games with pauses in between activities turned players collaborative |
| Malinverni et al. (2018) | Augmented & mixed reality impact children's understanding, collaboration and use of physical space | 36 | 9 -> 11 | 0 | Mixed-methods: • Between subjects experiment • Observations • Multimodal • Interviews | Not stated | "World-as-Support" portable projector system "Window-on-the-World" tablet app | Children played in groups of 4-5 using either a shared tablet OR a portable projector. Children were not instructed on how to share the device, they were allowed to organise the use of the device according to their preferences. | Players needed to find "magical portals" located on the walls of two different spaces within the school in order to solve a mystery. The magical portals provided relevant information to solve the mystery. | • recognition of physical space markers • projection of virtual content into physical space • display virtual content onto images of physical space | • Children's interactions with the system • Group interactions • Group collaboration • Use of physical space • Children's play experiences | • Children using the portable projector performed more gestures to express emotions, and more verbal exchanges focused on co-constructing knowledge with peers. They also physically arranged themselves in a semi-circle to work together • Children using the shared tablet established clear divisions of roles instead and their physical arrangement was more scattered • Children engaged in problem-solving to solve a mystery when using the projector or the tablet |
| Martin-Niedecken (2018) | Motion controllers impact on children's social and bodily interplay | 32 | 10 -> 14 | 0 | Mixed-methods: • Within subject experiment • Observations • Survey | 2 X 4 min sessions | "Plunder Planet" Kinect videogame Interactive projector | Participants played the videogame with an unacquainted peer twice. They stood side by side and shared one screen while using the two body motion controllers: FBMC or Kinect. | Players operated in the world of a young pirate, searching for buried treasures with a flying ship, collecting crystals, avoid collisions and fight off attacks of giant sandworms. | • adapt difficulty and complexity to the player's heart rate and in-game performance • FBM controller: virtual buttons at three physical heights • Kinect controller: operate through six levels using smart-eye technology | • Feeling of empathy, negative affect, social engagement, physical movements • Play behaviours: bodily interplay, communication, offensive/defensive play | • Both FBMC and Kinect facilitated some level of collaborative coordinated play • Participants engaged more in interdependent bodily interplay and communication with the FBMC than the Kinect • Participants without sports skills but with gaming skills interacted more confidently and were more cooperative when using FBMC |
| McKenzie et al. (2014) | Mobile-phone game impact on children's outdoor physical activity | 14 | 5 -> 12 | 0 | Quantitative • Survey • Analytics | 11 min sessions average | "Pirate Adventure" smartphone app | Children played the game outdoors individually with a smartphone. The smartphone would show the treasure map, clues and challenges. | Players hunt for treasures on a map with a set of predefined clues and movement activities challenges. Players followed paths and planned their route in pursuit of the next treasure. | • real world treasure location • player location tracking with respect to the game treasures • tracking of game status • display of game progress • tracking of physical activity | • Physical activity (PA) • Fundamental movement skills (FMS: hopping, side-stepping, jumping, running) | • In their self-reports almost all children (13) indicated they felt doing a lot of running and jumping during the game, and 5 of them considered themselves tired after it • Most children found the level of difficulty to be between "just right" and "too hard" • Information of player's relative position to other players made children go faster or take alternative routes between clues to beat competitors • Children engaged in deciding about the most strategic ways of advancing to the next clue location within a treasure-hunt game |
| Saksono et al. (2015) | Collaborative exergame impact on children's and parent's physical activity | 14 | ~8 | 15 | Qualitative • Interviews • Workshop | Sessions over 3 weeks | "Space Launched" exer-videogame computer app | Adults and children were left to use the app game along with the Fitbit activity tracking devices for various days. | Players engaged in physical activity to gain videogame time and unlock levels of the game to launch rockets towards planets, visit planets, observe launches, land on planets and complete space missions. | • tracking of physical activity time • physical activity activating videogame fuel • fuel progress display • inactivity reminder | • Salience of intention for physical activity • Adult-child (collaborative/competitive) interactions • Reasons driving interactive family game behaviours | • It increased awareness of opportunities for exercise and motivated those already aware of physical activity opportunities to follow them up • Caregivers and children collectively assessed their physical activity • Most participants preferred to use the game in competitive (comparing not outperforming each other) rather than collaborative ways |

| Author & year | Objective (To study:) | #Children | Age | #Adult | Methodology | Play Time | Play Tool | Study Play Setup | Play Goals | Key Play Tech Features | Variables studied | Main findings (relevant to review RQs) |
|---|---|---|---|---|---|---|---|---|---|---|---|---|
| Shahid et al. (2014) | Social robot impact on children's enjoyment and expression | 256 | 8 -> 12 | 0 | Quantitative • Between subjects experiment • Observations | 15 min sessions | "iCat" social robot Card game computer app | Children played the card game six times, either alone, in collaboration with a friend or with the iCat robot. | Players have to guess whether the next number of a sequence will be bigger or smaller than the previous (reference) number from the cards that come up in a deterministic computer game. When winning children collected coins as reward. | • robot waiting for child's suggestion to respond/react • robot to initiate interaction • robot to express both verbal and non-verbal behaviours | • Children's expressiveness of visual and social cues, positive and negative emotions • Behaviours showing connectedness, responsiveness, and compliance | • Younger children were more expressive while playing • Children playing with the social robot were more expressive than children playing alone, but less expressive than children playing with a friend • Children were more compliant with the social robot than with their friends • Children tended to seat closer to friends than the robot • Children engaged in problem solving to play the game * (children had to guess if an upcoming number was higher or lower than a previous number within a row of 6 cards) |
| Shahid (2018) | Video mediated communication system affect on children's feeling of social presence and socio-emotional response | 108 | ~ 8 | 0 | Quantitative: • Survey • Observations | 20 min sessions | A virtual card game played through video-conferencing platform | Game sessions at the childrens' school, each child played the card videogame 6 times. | Same as Shahid et al. (2014) above. | • two-player game • web-cam rendering of partner's face during joint remote play | • Fun and social experience: perceived co-presence, perceived message understanding, and perceived affect understanding | • Children's socio-emotional perceptions of co-presence as well as message and affect understanding were higher in the ideal and mutual gaze conditions • Children had least fun, were least expressive and did not feel a social bond in the no-gaze condition |
| Shen et al. (2018) | Social robot impact on children's interpersonal conflict and resolution skills | 64 | 3 -> 6 | 0 | Quantitative • Experiment | 5 X 10 sessions | "Keepon" social robot Traditional toys | Children played five activities in pairs across two conditions: 1) facilitated by Keepon robot; 2) facilitated, directed and mediated by the robot for social conflict. | Players played the games of choice: Lego Duplo ice cream set, magnetic tiles, toy house, remote control car, and making a birthday card with crayons and stickers. | • robot facilitated play • voice-enabled activity interaction • conflict detection via sound • prompts to solve conflicts • encouragement via sounds | • The socialness of play behaviours • The constructiveness of play behaviours • Object possession conflict | • There were no differences in socialness, constructiveness or amount of conflicts across conditions • When the social robot mediated conflicts, children were 4 times more likely to resolve conflicts in positive ways • The pausing of play enforced by the social robot following a social conflict helped children resolve their conflicts |
| Straker et al. (2009) | Comparison between different types of screen interaction and impact on children's muscle activity | 20 | 9 -> 12 | 0 | Quantitative • Within subject experiment | 1.5 hour sessions | Animated TV film Different computer videogames | Children played different videogames in the same order using different types of controllers. Multiple sensors were worn by the child to capture muscular activity. | Children watched "The Incredibles" animated film or played the videogames of choice. | • videogame operated via an array of controllers: console, gamepad, keyboard, steering wheel, pedals, EyeToy device | • Movement and muscular activity in various parts of the body | • Use of the wheel controller resulted in some increase in upper limb movement and muscle activity, but the other traditional input devices were usually as sedentary as watching a DVD • Use of the active-input device based on body movements (EyeToy) resulted in considerably greater activity at all muscles |

* Results information inferred by reviewers from the report of each study